\documentclass[bibyear]{aa}  
\usepackage{graphicx,amssymb}
\usepackage{color}
\usepackage[varg]{txfonts}
\usepackage{url}%  kod po poprawkach: dobry_kod.f

\begin{document}

   \title{SALT long-slit spectroscopy of quasar HE 0435-4312:\\fast displacement of the  Mg II emission line \thanks{Based on observations made with the Southern African Large Telescope  (SALT) under program 2012-2-POL-003 and 2013-1-POL-RSA-002  (PI: B. Czerny). Spectra shown in Figs.~3 and 4 are only available in electronic form at the CDS (Centre de Données astronomiques de Strassbourg) via anonymous ftp to cdsarc.u-strasbg.fr (130.79.128.5)
or via http://cdsweb.u-strasbg.fr/cgi-bin/qcat?J/A+A/.}}

   \author{J. \' Sredzi\' nska\inst{1} \and B. Czerny\inst{1,2}  \and K. Hryniewicz\inst{1}  \and  M. Krupa\inst{3} \and A. Kurcz\inst{3} \and P. Marziani\inst{4} \and \\ T. P. Adhikari\inst{1} 
   \and R. Basak\inst{1} \and B. You\inst{1} \and J.-M. Wang\inst{5} \and C. Hu\inst{5} \and W. Pych\inst{1} \and M. Bilicki\inst{6}}
   \institute{Nicolaus Copernicus Astronomical Center, Polish Academy of Sciences, Bartycka 18, 00-716 Warsaw, Poland 
\and
Center for Theoretical Physics, Polish Academy of Sciences, Al. Lotnikow 32/46, 02-668 Warsaw, Poland
\and
 Astronomical Observatory of the Jagiellonian University, Orla 171, 30-244 Cracow, Poland
\and
INAF, Osservatorio Astronomico di Padova, 35122 Padova, Italy
\and
Key Laboratory for Particle Astrophysics, Institute of High Energy Physics, Chinese Academy of Sciences, 19B Yuquan Road,
Beijing 100049, China
\and
Leiden Observatory, Leiden University, Niels Bohrweg 2, NL-2333 CA Leiden, the Netherlands
%\and 
%Warsaw University Observatory, Al. Ujazdowskie 4, 00-478 Warszawa, Poland
             }

   \date{Received ............; accepted ..............}

% \abstract{}{}{}{}{} 
% 5 {} token are mandatory
 
  \abstract
   {The Mg~II emission line is visible in the optical band for intermediate redshift quasars ($0.4<z<1.6$) and it is thus an extremely important tool to measure the black hole mass and to understand the structure of the Broad Line Region.}
   { We aim to determine the substructure and the variability of the Mg~II line with the aim to identify which part of the line comes from a medium in Keplerian motion.}
   {Using the Southern African Large Telescope (SALT) with the Robert Stobie Spectrograph (RSS) we performed ten spectroscopic observations of quasar HE 0435-4312 ($z=1.2231$) over a period of three years (Dec 23/24, 2012 to Dec 7/8, 2015).}
   {Both the Mg II line and the Fe II pseudo-continuum increase with time. We clearly detect the systematic shift of the Mg II line with respect to the Fe II over the years, corresponding to the acceleration of $104 \pm 14$ km s$^{-1}$ year$^{-1}$ in the quasar rest frame. The Mg II line shape is clearly non-Gaussian but single-component, and the increase in line equivalent width and line shift is not accompanied with significant evolution of the line shape. We analyse the conditions in the Mg II and Fe II formation region and we note that the very large difference in the covering factor and the turbulent velocity also support the conclusion that the two regions are spatially separated.}
   {The measured acceleration of the line systematic shift is too large to connect it with the orbital motion at a distance of the Broad Line Region (BLR) in this source. It may imply a precessing inner disk illuminating the BLR. Further monitoring is still needed to better constrain the variability mechanism.}

\authorrunning{\' Sredzi\' nska et al.}
\titlerunning{SALT long-slit spectroscopy of HE 0435-4312}

   \maketitle
%
%________________________________________________________________

\section{Introduction}

Quasars  represent a very important subclass of active galactic nuclei (AGN). These objects are highly luminous and detected in a wide range of redshifts. Central parts of quasars are currently 
unresolved, but studies of the broad emission lines originating from the Broad Line Region (BLR) located around 1 pc from the nuclei give insight into the gas dynamics, accretion inflow and outflows, and provide estimates of black hole (BH) masses.

A few broad permitted lines dominate AGN spectra, and they are divided into Low Ionization Lines (LIL) and  High Ionization Lines (HIL; Collin-Souffrin et al. 1988). LILs, such as Balmer lines, Mg II, C II, and Fe II originate from a high density medium (local density of the order of $10^{11}$ cm$^{-3}$ or above) and most likely are generated close to the accretion disk surface (Collin-Souffrin et al. 1988; Matsuoka et al. 2008; Negrete et al. 2012; Ruff et al. 2012; Zhang 2013; Martinez-Aldama et al. 2015, Schnorr-M\" uller et al. 2016) while HILs, such as Ly$\alpha$, C III, C IV, He I, He II, and N V  originate from a lower density medium (local density of the order of $10^9$cm$^{-3}$ or less) (Collin-Souffrin et al. 1988).  LILs are more symmetric,
so the motion of the emitting gas is mostly Keplerian, but HILs show more of a blueshift, implying 
outflow (Gaskell
1982; Wilkes 1984, Wills et al. 1993; Corbin and Boroson 1996; Marziani et al. 1996; Baldwin et al. 1996; Richards et al.  2002; Shin et al. 2016; Veilleux et al. 2016). Thus LILs are preferred for the determination of black hole mass, but for high redshift quasars HIL C IV offers the only option if spectra in the optical band are available (Vestergaard \& Peterson 2006; Kelly et al. 2010).

Line intensity varies in response to the variable emission from the innermost parts of the nucleus as originally discovered by Andrillat \& Souffrin (1968). This fundamental result was later confirmed in countless studies of nearby AGN, as well as of distant quasars (e.g., Kaspi et al. 2007; Woo 2008; Woo et al. 2013, Cackett et al. 2015). BLR reverberation studies (i.e., measurements of the time delays of the line with respect to the continuum), done later for numerous, mostly nearby, objects  (e.g., Cherepashchuk \& Lyutyi 1973; Gaskell \& Sparke 1986; Wandel et al. 1999, Kaspi et al. 2000, Metzroth et al. 2006; Bentz et al. 2013; for a recent compilation see e.g. Du et al. 2015, 2016) showed delays from a few hours (NGC 4395; Peterson et al. 2005) to 300 days (Kaspi et al. 2000), and perhaps longer for high redshift objects (Kaspi et al. 2007). In general, the delays for LILs were longer than for HILs ( e.g., Wandel et al. 1999; Desroches et al. 2006; Clavel et al. 1991; Reichert et al. 1994; Korista et al. 1995; Goad et al. 2016).

The most studied LIL is H$\beta$ but for higher redshift sources it moves to the infrared (IR) while Mg II moves into the optical band (for redshifts between 0.4 and 1.5). Studies of the sources with both H$\beta$ and Mg II in the optical band showed similar properties of the two lines and the viability of both for black hole mass measurement (e.g., Kong et al. 2006; Shen et al. 2008; Vestergaard \& Osmer 2009), possibly with a minor rescaling (Wang et al. 2009, Shen \& Liu 2012; Trakhtenbrot \& Netzer 2012, Marziani et al. 2013ab). Those studies, statistical in their character and based on a single spectrum for a given object, do not address, however, the issue of the variability of the line profile.  

Intrinsic variations of the emission line profile (i.e., not just the line intensity) have been measured for a few sources, almost always for Balmer lines (e.g., NGC 4151: Shapovalova et al. 2008, 2010a, 3C 390.3: Gaskell 1996, Shapovalova et al. 2010b; Arp 102B: Shapovalova et al. 2013, Popovic et al. 2014; Ark 564: Shapovalova et al. 2012; NGC 1097: Storchi-Bergmann et al. 1995, Schimoia et al. 2015;  NGC 5548: Peterson et al. 1987, Shapovalova et al. 2004, Sergeev et al. 2007, Bon et al. 2016; see Ilic et al. 2015 for a campaign summary). Variability of the Mg II and C IV line wings was explored by Punsly (2013), but in this case part of the emission is related to the radio jet mechanism. Numerous examples of line broadening or weakening with the change of the flux were discovered through systematic searches for changing-look AGNs (MacLeod et al. 2016; Runco et al. 2016). The change in the H$\beta$ line position with respect to the systemic redshift in two-epoch spectroscopy has been detected for several quasars (Shen et al. 2013; Liu et al. 2014). Those variations are frequently interpreted as a signature of a binary black hole at the nucleus (e.g., Gaskell 1983; Peterson et al. 1987; Gaskell 1996; Li et al. 2016; but see Leighly et al. 2016 for arguments against this view for Mkn 231).  Such a kind of variation is different from line variations due to absorption (e.g., Hall et al. 2011; Misawa et al. 2014; Wildy et al. 2014; De Cicco et al 2016).  As Shen et al. (2013) and Liu et al. (2014) showed, quasar intrinsic line shape variations are statistically rare, the changes were detected only in 4\% of quasars, and the study of the line profile changes was never done for the Mg II line although the variability of this line intensity was firmly established (e.g., Clavel et al.
1991; Reichert et al. 1994; Trevese et al. 2007;  Woo et al. 2008; Sun et al. 2015; Cackett et al. 2015; Hryniewicz et al. 2014). In this paper we study possible intrinsic variations of the shape of the Mg II line in the quasar HE 0435-4312 from ten observations done with the Southern African Large Telescope (SALT) telescope. We also analyze the broad band data and put constraints on the conditions within the BLR.

\section{Observations}

In the present paper we concentrate on the quasar HE 0435-4312 ($z = 1.232$, $V = 17.1$~mag, coordinates for the epoch J2000: RA = 04h37m11.8s, DEC = -43d06m04s, as given in NED\footnote{NASA/IPAC Extragalactic Database (NED) is operated by the Jet Propulsion Laboratory, 
California Institute of Technology, http://ned.ipac.caltech.edu/.}). The source was discovered in the course of the Hamburg 
quasar survey\footnote{\url{http://www.hs.uni-hamburg.de/EN/For/Exg/Sur/hes/qso_surveys.html}.} (Wisotzki et al. 2000). We selected it, together with two other quasars, for reverberation studies of bright intermediate redshift quasars with the Southern African Large Telescope (SALT) (Czerny et al. 2013, Hryniewicz et al. 2014, Modzelewska et al. 2014).

\subsection{Spectroscopy}
We performed ten observations in the optical band of the object HE 0435-4312 with the Robert Stobie Spectrograph 
(RSS; Burgh et al. 2003, Kobulnicky et al. 2003; Smith et al. 2006) on the SALT telescope, in the service mode. The data were collected between  Dec 23/24 2012, and Dec 7/8 2015 (see Table~\ref{table:daty}), thus they cover the period of almost exactly three years. 
Every observing block contained two $\sim 12$ minute exposures, in a long slit mode, with the slit width of 2", 
an exposure of the calibration lamp, and a number of flat-field images. The CCD detector of the RSS consists of a mosaic of three CCD matrices, a total size of $6362 \times 4102$ pixels, with a single pixel size of 15 $\mu$m, corresponding to a spatial resolution of 0.1267 arc seconds per pixel. We used the $2 \times 2$ binning readout option which reduced the resolution by a factor of two but increased the signal to noise (S/N) ratio. The mean gain of the mosaic is 1.7, the readout noise $\sim 3.5$ electrons.  
 We used RSS PG1300 grating, and the grating tilt angle 26.75 deg which gives the observed wavelength coverage from 5897 \AA ~ to 7880 \AA. The average resolving power of this configuration is 1420. The spectral resolution  at 5500 \AA~ is $R=1047$. Order blocking was done with the blue PC04600 filter.

\begin{table*}
\caption{Log of spectroscopic observations with SALT.}   % title of Table
\label{table:daty}      % is used to refer this table in the text
\centering                          % used for centering table
\begin{tabular}{l r r r r r }        % centered columns (4 columns)
\hline\hline      % inserts double horizontal lines
no    & date & MJD  & exposure &  seeing  & comments\\
      &      &  -2450000    &   [s]    &  `` & \\
\hline
1     &  23~Dec~2012 & 6284.50& $2 \times 739.2$ & 1.4 & photometric, no clouds, dark\\
2     &  18~Feb~2013 & 6341.50& $2 \times 739.2$ & 1.1 & photometric, clear, gray to dark\\
3     &  20~Aug~2013 & 6524.50& $2 \times 618.2$ & 1.6-1.7 & photometric, clear, bright\\
4     &   3~Feb~2014 & 6691.50& $2 \times 667.2$ & 1.6 & photometric, no clouds, clear, dark\\
5     &  23~Aug~2014 & 6892.50& $2 \times 747.2$ & 1.7-1.8 & not photometric, dark\\
6     &  24~Nov~2014 & 6985.50& 820.2, 390.8     & 1.5 & not photometric, clouds\\
7     &   9~Jan~2015 & 7031.50& $2 \times 820.2$ & 3.4 & photometric, clear, gray\\
8     &   9~Aug~2015 & 7243.50& $2 \times 795.2$ & 1.7 & photometric, moon 21\% and low in sky\\
9     &   6~Oct~2015 & 7301.50& $2 \times 795.2$ & - & no information about night, moon 29\%\\
10    &   7~Dec~2015 & 7363.50& $2 \times 823.2$ & 1.2 & mostly clear, moon 11\%\\
\hline                                   %inserts single line
\end{tabular}
\end{table*}  

Most observations were performed on photometric nights, without intervening clouds and in dark and/or gray moon conditions, with seeing $\sim 1.5$", but in some observations thin clouds were present 
(for details, see Table~\ref{table:daty}). No information about the observing conditions was provided for observation 9. 

The preliminary data reduction (gain correction, overscan bias subtraction, cross-talk correction and amplifier mosaicing) was  
done with a semi-automated pipeline from the SALT PyRAF package\footnote{http://pysalt.salt.ac.za.} (see Crawford et al. 2010)
by SALT observatory staff. We performed further analysis (including flat-field correction) 
using the IRAF package\footnote{IRAF is distributed by the National Optical Astronomy Observatories, which are operated
by the Association of Universities for Research in Astronomy, Inc., under cooperative agreement with the NSF.} 
%(dataio, noao.imred.ccdred, images.imutil, images.imfit,
%images.immatch, noao.twodspec.longslit, noao.twodspec.apextract)
. 
Pairs of exposures were combined into a single image to increase the S/N, and to remove cosmic ray effects. We calibrated the spectrum using lamp exposures 
(different lamps were used on different nights), and finally we extracted one-dimensional spectra with the IRAF noao.twodspec package. We checked the lamp calibration against the sky lines 
seen in the background files. The OI$\lambda 6300.304$ \AA ~ line is reliable (Osterbrock et al. 2000) and well visible in all ten observations. 
Observation 1 clearly needed a considerable shift by $\sim 7.5$ \AA~ in order to move the sky line O I into the appropriate position. We therefore estimated the requested shift with respect to the reference provided by O I line
for all ten data sets. Since the O I was not perfectly symmetric, we used the median value to determine the observed position of O I, and the difference between the median and the expected wavelength was used to shift the whole observed spectra. Apart from Observation 1, the shifts are much smaller than the pixel size of 0.62 \AA. We provide these values in Table~\ref{tab:shift}. 

\begin{table}
\caption{Recalibration of the SALT spectra at the basis of the background OI$\lambda 6300.304$ \AA~ line}   % title of Table
\label{tab:shift}      % is used to refer this table in the text
\centering                          % used for centering table
\begin{tabular}{l r r r r r }        % centered columns (4 columns)
\hline\hline      % inserts double horizontal lines
Observation    & Shift\\
      & [\AA]   \\
\hline
1     &  7.40  \\
2     &  0.00\\
3     &  -0.34\\
4     &  -0.15\\
5     &  0.20\\
6     &  -0.13\\
7     &  -0.27\\
8     &  0.22\\
9     &  -0.20\\
10    &  -0.12\\
\hline                                   %inserts single line
\end{tabular}
\end{table}

Since the SALT telescope has serious vignetting problems, we further corrected the overall spectral shape following the general method described in Modzelewska et al. (2014). We obtained the 
correcting function using the star LTT 4364, which was observed with SALT on February 18, 2013, with the instrumental setup PG1300/26.7500 (spectrum P201302170120), and was also available at 
the ESO website.\footnote{ftp://ftp.eso.org/pub/stecf/standards/ctiostan/.} This was optimized to accurately model the 5000 - 5700 \AA~  wavelength range. In further analysis we neglected the 
instrumental broadening since it is unimportant for SALT quasar broad emission lines (Hryniewicz et al. 2014).

Next we de-reddened the spectra although the Galactic extinction in the direction of  HE 0435-4312 is very low
 ($A_{\lambda} = 0.060, 0.045$, and 0.036 in the B,V, and R bands respectively; Schlafly \& Finkbeiner 2011 after NASA/IPAC Extragalactic Database (NED)). 
We did not consider the intrinsic absorption in the source. We further assumed that the host galaxy does not contribute to the
UV part of the quasar spectrum and we modeled the observed spectrum as coming directly from the active nucleus.  

\subsection{Photometry}

\begin{figure}[ht]
    \centering
   %\includegraphics[width=0.95\hsize]{lightcurve_V.eps}
   %\includegraphics[width=0.95\hsize]{lightcurve_v2.eps}
    %   \vskip - 0.5 true cm
    \includegraphics[width=0.49\textwidth]{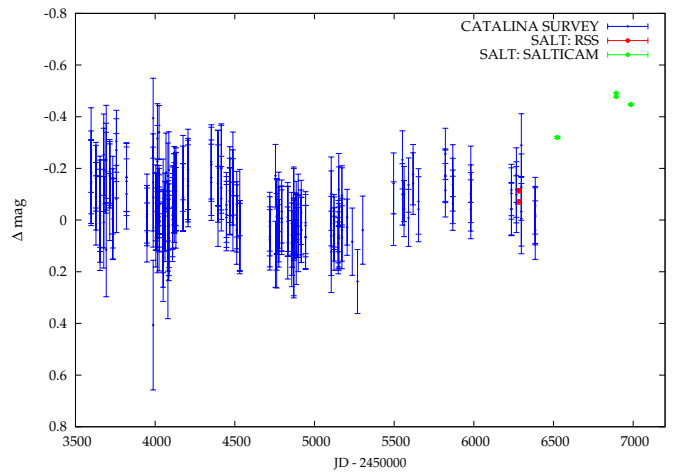}
    %\vskip - 1.5 true cm
  %  \includegraphics[width=0.5\textwidth]{lightcurve_V_long_v3.eps}
     %  \vskip - 1.0 true cm
    \caption{Catalina light curve for quasar HE 0435-4312 (blue points). Red points mark the SALT light curve from RSS Clear Imaging mode, and green points are from SALTICAM V band, and SALTICAM g-band, all of them arbitrarily shifted as a set to match the Catalina points.}
               \label{fig:lightcurve}%
    \end{figure}

We supplemented the spectroscopic data with quite long coverage but not very accurate photometry (typical uncertainty of 0.15 mag; or less as argued by Vaughan et al. 2016)
from the  Catalina survey (Drake et al. 2009\footnote{\url{http://www.lpl.arizona.edu/css}.}) which covers eight years 
(see Fig.~\ref{fig:lightcurve}). This period partially overlaps with our spectroscopic data. We can see the continuum variability, at a level of $\sim 20$ \%, 
in this timescale. 

In principle SALT allows us to make photometric measurements using SALTICAM (UV-Visible 320 - 950 nm imaging and acquisition camera). We attempted such measurements whenever the SALTICAM data were available for our object, and in one case we could use RSS in Clear Imaging mode. Those photometric points are not well calibrated since they came from different filters. Results are overplotted in Fig.~\ref{fig:lightcurve}, with a shift making the two measurement sets coincide in the overlap region.  We do not use these data in further analysis but only treat them as a confirmation of the quasar variability.

%______________________________________________ 
%
%1) SALT_2012_obs3: RSS - nie ma filtra
%2) SALT_2013_obs4: SALTICAM - filtr V-S1
%3) SALT_2014_obs1: SALTICAM - filtr SDSSg-S1
%4) SALT_2014_obs3: SALTICAM - filtr SDSSg-S1
%___________________________

\subsection{Broad band spectrum}
\label{broad_band_spectrum}

In order to discuss the physical conditions in the quasar BLR we need to know the broad band continuum irradiating the clouds. We also use the modeling of the broad band spectrum with an accretion disk model to obtain the constraints on the black hole mass, and to test the single or multiple component character of the Mg II line using the relationship between the line width, quasar luminosity, and the black hole mass (e.g., Bentz et al. 2013).

\begin{figure}
   \centering
\includegraphics[width=0.5\textwidth]{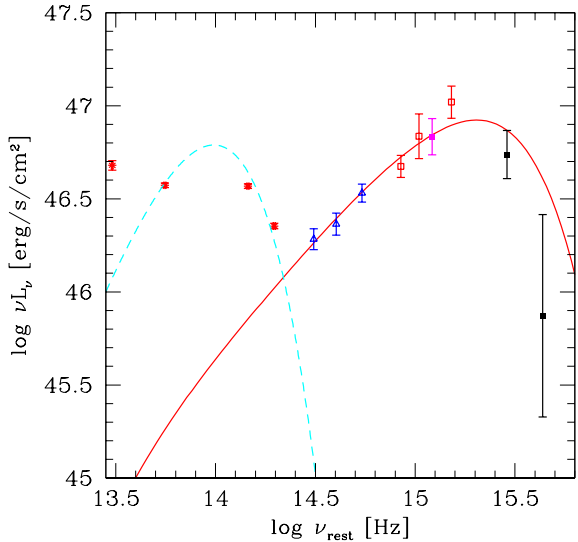}
   \caption{Time-averaged broad band spectrum of HE 0435-4312 in the IR and optical band (points) in the rest frame (red stars - WISE, blue triangles - 2MASS, red open squares - USNO, filled magenta square - mean Catalina level, black squares - GALEX), together with the parametric fit by the disk (see Sect.~\ref{sect:BHmass}) with the black hole mass $M = 2.2 \times  10^9 M_{\odot}$, accretion rate $\dot m = 0.58$ in Eddington units, viewing angle $i = 23^{\circ}$, and non-rotating black hole spin $a = 0$. The mid-IR  contribution likely comes from the  circumnuclear dust and it was modeled as a black body of a temperature of 1100 K (cyan dashed line). Data points are not simultaneous.}
              \label{fig:broad_band}%
    \end{figure}

The quasar HE 0435-4312 has been observed at various wavelengths during the last ten years. We collected non-simultaneous data with the use of the ASDC (ASI Science Data Center) 
Sky Explorer/SED Builder web interface\footnote{http://tools.asdc.asi.it/.} and GALEX View.\footnote{http://galex.stsci.edu/GalexView/.} The data points come from WISE (Wide-field Infrared Survey Explorer; Wright et al. 2010), 2MASS (Two Micrron All-Sky Survey; Skrutskie et al. 2006),  USNO (The United State Naval Observatory) A2.0, USNO B1, and GALEX (Galaxy Evolution Explorer; Martin et al. 2005). 
%and RASS (Voges et al. 1999). 
The combined spectrum was corrected for the Galactic extinction ($E(B-V) = 0.015$ from NED) using the Cardelli et al. (1989) extinction curve. The result is shown in Fig.~\ref{fig:broad_band}. 
The highest flux point at $\log \nu = 14.736$ comes from a very old survey USNO, and the exact date of this observation is unknown. A data point from this survey was also problematic for the previously 
studied source CTS C30.10 (Modzelewska et al. 2014).  
The observed fluxes are converted to the absolute values assuming the cosmological 
parameters: $H = 71$ km s$^{-1}$ Mpc$^{-1}$, $\Omega_{\lambda} = 0.73$, $\Omega_M = 0.27$.

%---------------------------------------------------------------------

%______________________________________________  
%
%
%______________________________________________________________

\begin{table*}
\caption{Fits of a single Gaussian (G), single Lorentzian (L), Edgeworth expansion (E),  Gauss-Hermite expansion (GH),  and double Lorentzian (LL) shape with an arbitrary redshift for the Fe II template 1 (marked with $^*$), or template 3, for ten individual spectra obtained
with SALT between December 2012 and December 2015.}   % title of Table
\label{tab:free_redshift}      % is used to refer this table in the text
\centering                          % used for centering table
\begin{tabular}{l r r r r r r r r r r}        % centered columns (4 columns)
\hline\hline      % inserts double horizontal lines
Obs. & G & G & L &L  & E & E & GH & GH & LL & LL \\
               &  redshift    &  $\chi^2$   &  redshift    &  $\chi^2$  &  redshift    &  $\chi^2$  & redshift & $\chi^2$ & redshift & $\chi^2$\\
\hline
1  &  1.22190 & 725.5   &  1.22189 & 811.7 & 1.22311 & 670.5  &1.22311 &  667.6 & 1.22312 & 635.0\\
2  &  1.22225 & 1360.4  &  1.22224 &1371.9 & 1.22306 & 1246.3 &1.22306 & 1197.9 & 1.22306 & 1237.5\\
3  &  1.22237 & 513.5   &  1.22237 & 628.6 & 1.22321 & 509.2  &1.22320 &  505.3$^*$& 1.22320 & 498.3\\
4  &  1.22237 & 595.8   &  1.22236 & 696.9 & 1.22319 & 557.9  &1.22319 &  541.6$^*$& 1.22319 & 531.3\\
5  &  1.22237 & 711.9   &  1.22237 & 842.9 & 1.22307 & 698.8  &1.22307 &  675.0$^*$& 1.22296 & 645.7\\
6  &  1.22246 &1267.6   &  1.22246 &1192.0 & 1.22303 & 1251.3 &1.22302 & 1220.6$^*$& 1.22302 & 1208.3\\
7  &  1.22245 & 587.5   &  1.22246 & 633.0 & 1.22302 & 547.5  &1.22302 &  538.7$^*$& 1.22302 & 511.8\\
8  &  1.22236 & 1408.3  &  1.22237 &1535.3 & 1.22311 & 1404.8 &1.22320 &  1400.0   & 1.22320 & 1434.1\\
9  &  1.22215 & 903.12  &  1.22215 & 999.2 & 1.22304 & 900.6  &1.22299 &   860.1   & 1.22300 & 853.2\\
10 &  1.22235 & 614.7   &  1.22234 & 697.8 & 1.22315 & 536.9  &1.22231 &  515.6$^*$& 1.22312 & 520.9\\
\hline                                   %inserts single line
\end{tabular}
\end{table*}

%______________________________________________ 
%
    \begin{figure}[ht]
    \centering
    %   \vskip - 0.5 true cm
    \includegraphics[width=0.45\textwidth]{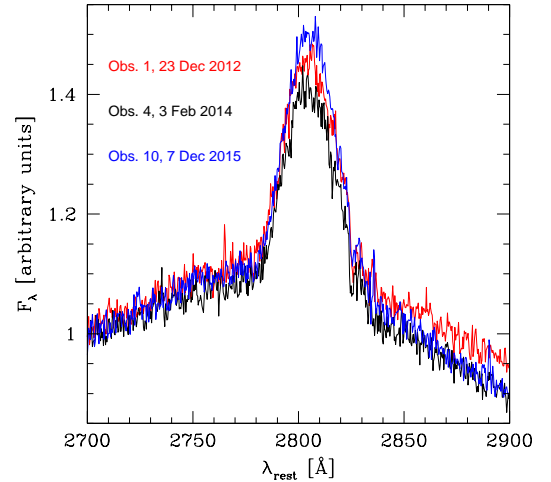}
    %\vskip - 1.5 true cm
    \caption{Three examples of quasar spectra in the Mg II region, renormalized to the 2700 \AA~flux, clearly show variations of the Mg II line, but accompanied with the variability of the underlying Fe II and the power law slope.}
               \label{fig:widma_sample}%
    \end{figure}
%
%___________________________

%___________________________

\begin{table*}
\caption{Fits of the single Gaussian shape of Mg II  with errors (90 \% confidence level) for template 3 (broadening 
900 km s$^{-1}$) for ten individual spectra obtained with SALT
between December 2012 and December 2015.}   % title of Table
\label{tab:SG}      % is used to refer this table in the text
\centering                          % used for centering table
\begin{tabular}{l r r r r r r r r r}        % centered columns (4 columns)
\hline\hline      % inserts double horizontal lines
Obs.    & Mg II                    &  Mg II                     & Mg II                                  &  Fe II          \\
        & EW                       &  $\sigma$                  & s                                      &  EW             \\
        & \AA                      &  km s$^{-1}$               & $10^{-3}$                               &  \AA            \\
\hline                        % inserts single horizontal line
1       &  $19.38^{+0.27}_{-0.26}$ & $1539^{+17}_{-17}$    & $2.25^{+0.05}_{-0.05}$ & $24.33^{+1.09}_{-1.03}$ \\
2       &  $18.61^{+0.28}_{-0.30}$ & $1510^{+18}_{-18}$    & $2.54^{+0.05}_{-0.05}$ & $24.25^{+1.14}_{-1.09}$ \\
3       &  $20.25^{+0.49}_{-0.44}$ & $1530^{+26}_{-25}$    & $2.37^{+0.08}_{-0.08}$ & $26.52^{+1.88}_{-1.85}$ \\
4       &  $19.15^{+0.34}_{-0.31}$ & $1518^{+19}_{-19}$    & $2.42^{+0.05}_{-0.05}$ & $24.27^{+1.18}_{-1.20}$ \\
5       &  $21.31^{+0.31}_{-0.32}$ & $1545^{+18}_{-18}$    & $2.61^{+0.05}_{-0.06}$ & $27.97^{+1.28}_{-1.21}$ \\
6       &  $21.83^{+0.56}_{-0.59}$ & $1482^{+31}_{-29}$    & $2.67^{+0.09}_{-0.08}$ & $23.69^{+2.14}_{-2.07}$ \\
7       &  $21.80^{+0.45}_{-0.47}$ & $1527^{+23}_{-23}$    & $2.63^{+0.07}_{-0.07}$ & $27.11^{+1.70}_{-1.72}$ \\
8       &  $20.83^{+0.28}_{-0.29}$ & $1532^{+16}_{-16}$    & $2.58^{+0.04}_{-0.05}$ & $27.12^{+1.16}_{-1.06}$ \\
9       &  $21.31^{+0.33}_{-0.36}$ & $1560^{+19}_{-19}$    & $2.77^{+0.06}_{-0.06}$ & $28.52^{+1.40}_{-1.36}$ \\
10      &  $23.01^{+0.30}_{-0.28}$ & $1517^{+16}_{-16}$    & $2.62^{+0.05}_{-0.04}$ & $28.99^{+1.14}_{-1.12}$ \\
\hline                                   %inserts single line

\end{tabular}
\end{table*}

%___________________________

\begin{table*}
\caption{Fits of the Edgeworth (E) and Gauss-Hermite (GH) expansion for the shape of Mg II  with errors (90 \% confidence level) for the best template choice (template 3, broadening 1550 km s$^{-1}$ for E and template 1, broadening 1400 km s$^{-1}$ for GH), for ten individual spectra obtained with SALT
between December 2012 and December 2015.}   % title of Table
\label{tab:final_Hermit}      % is used to refer this table in the text
\centering                          % used for centering table
\begin{tabular}{l r r r r r r r r }        % centered columns (4 columns)
\hline\hline      % inserts double horizontal lines
Obs.    & Mg II~~(E)                    &  Mg II~~(E)                     & Mg II~(E)                   &  Fe II ~(E)  &  Mg II~~(GH)                    &  Mg II~~(GH)                     & Mg II~(GH)                   &  Fe II ~(GH)            \\
        & EW                       & $\lambda$3                      & $\lambda$4                      &  EW  & EW                       & h3                      & h4                      &  EW             \\
        & \AA                         &             &                      &  \AA   & \AA                         &             &                      &  \AA            \\
\hline                        % inserts single horizontal line
1       &  $20.17^{+0.31}_{-0.34}$ &  $0.28^{+0.09}_{-0.09}$ & $1.12^{+0.24}_{-0.22}$ & $22.99^{+1.09}_{-2.33}$ &$31.38^{+0.98}_{-1.02}$&$-0.125^{+0.006}_{-0.006}$& $0.10^{+0.01}_{-0.01}$&$26.5^{+2.3}_{-2.3}$\\
2       &  $19.30^{+0.28}_{-0.30}$ &  $0.30^{+0.10}_{-0.10}$ & $1.57^{+0.19}_{-0.19}$ & $19.13^{+1.39}_{-1.12}$ &$29.04^{+0.88}_{-0.85}$&$-0.131^{+0.008}_{-0.008}$& $0.12^{+0.01}_{-0.01}$&$23.0^{+2.4}_{-2.2}$\\
3       &  $20.73^{+0.53}_{-0.40}$ & $0.31^{+0.12}_{-0.14}$  & $0.99^{+0.32}_{-0.34}$ & $20.19^{+2.10}_{-2.76}$ &$32.09^{+1.64}_{-1.60}$&$-0.127^{+0.012}_{-0.010}$& $0.11^{+0.02}_{-0.02}$&$25.5^{+4.0}_{-3.9}$\\
4       &  $19.41^{+0.34}_{-0.30}$ &  $0.34^{+0.08}_{-0.09}$ & $0.85^{+0.32}_{-0.25}$ & $23.38^{+1.34}_{-1.90}$ &$31.89^{+1.17}_{-1.18}$&$-0.131^{+0.007}_{-0.007}$& $0.09^{+0.01}_{-0.01}$&$28.5^{+2.7}_{-2.8}$\\
5       &  $22.33^{+0.38}_{-0.28}$ & $0.47^{+0.06}_{-0.06}$  & $0.69^{+0.21}_{-0.18}$ & $27.36^{+1.59}_{-1.73}$ &$34.45^{+1.13}_{-1.15}$&$-0.122^{+0.007}_{-0.007}$& $0.12^{+0.01}_{-0.01}$& $27.9^{+2.7}_{-2.9}$\\
6       &  $22.48^{+0.64}_{-0.55}$ & $0.45^{+0.14}_{-0.13}$  & $0.90^{+0.33}_{-0.33}$ & $22.14^{+2.59}_{-2.80}$ &$33.86^{+1.91}_{-1.70}$&$-0.102^{+0.016}_{-0.011}$& $0.12^{+0.02}_{-0.02}$ & $25.0^{^+4.2}_{-4.1}$\\
7       &  $22.64^{+0.46}_{-0.49}$ &  $0.39^{+0.11}_{-0.11}$ & $1.08^{+0.27}_{-0.25}$ & $23.92^{+2.71}_{-2.00}$ &$34.53^{+1.55}_{-1.53}$&$-0.124^{+0.009}_{-0.010}$& $0.12^{+0.01}_{-0.01}$ & $27.8^{+3.7}_{-3.6}$\\
8       &  $21.97^{+0.34}_{-0.31}$ &  $0.54^{+0.07}_{-0.08}$ & $1.28^{+0.17}_{-0.16}$ & $23.33^{+1.26}_{-1.59}$ &$32.57^{+1.60}_{-1.60}$&$-0.116^{+0.007}_{-0.006}$& $0.12^{+0.01}_{-0.01}$ & $26.4^{+2.6}_{-2.5}$\\
9       &  $22.11^{+0.57}_{-0.33}$ &  $0.28^{+0.09}_{-0.11}$ & $0.98^{+0.24}_{-0.22}$ & $24.71^{+3.08}_{-2.65}$ &$32.81^{+1.24}_{-1.27}$&$-0.129^{+0.008}_{-0.009}$& $0.11^{+0.01}_{-0.01}$ & $24.0^{+3.0}_{-3.0}$\\
10      &  $23.96^{+0.38}_{-0.27}$ &  $0.37^{+0.06}_{-0.07}$ & $1.00^{+0.20}_{-0.18}$ & $26.44^{+1.63}_{-1.97}$ &$37.08^{+1.16}_{-1.10}$&$-0.126^{+0.006}_{-0.006}$& $0.11^{+0.01}_{-0.01}$ & $30.2^{+2.4}_{-2.5}$\\
\hline                                   %inserts single line

\end{tabular}
\end{table*}

\section{Model}

Analyzing SALT data we concentrate on modeling a relatively narrow spectral band, between 2700 \AA~and 2900 \AA~in the rest frame. We
assume that the spectrum consists of three basic components: power-law continuum, representing the emission of the accretion disk, Fe II pseudo-continuum, and the Mg II line. 

The Mg II line is treated as a doublet (2796.35 - 2803.53 \AA; Morton 1991). 
In most fits we assume the doublet ratio is 1:1, since in previously analyzed quasars this ratio was not 
well constrained (Hryniewicz et al. 2014; Modzelewska et al. 2014) as the line is unresolved. We test the sensitivity of the spectral fitting and the possibility to get constraints on the doublet ratio for some model subsets. The position
of the line is a free parameter since the previous determination of the redshift was based on low quality data (Maza et al. 1993). 

The kinematic shape of each of the doublet components is modeled either as a single Lorentzian, single Gaussian, double Lorentzian, double Gaussian, Edgeworth expansion (see, e.g., Blinnikov \& Moessner 1998), or  Gauss-Hermite expansion (see, e.g., Van der Marel \& Franx 1993; La Mura et al. 2009). The Fe II component is assumed to be a reference in the redshift determination, while the line emission is allowed to have a shift $s = V/c$ with respect to the systemic velocity. We allow such a shift also when using the Edgeworth or Gauss-Hermite expansion, and we limit ourselves to three terms, that is, up to the skewness and kurtosis terms only (see Appendix A).

The use of the Fe II as the redshift reference is a technically motivated choice since the Fe II component preparation requires the convolution of the basic template with the Fe II velocity dispersion. We do not detect significant narrow lines which could serve as a reliable absolute reference. However, a shift between the Mg II and  Fe II is already observationally established (Ferland et al. 2009).

If two Gaussian or two Lorentzian components are used, one of them is fixed at the position of the Fe II contribution while the other is allowed to have an arbitrary location. This reduced the number of the free parameters, and at the same time it is supported by two-component character of H$\beta$ fits, with one of the components kinematically similar to the optical Fe II emission (Hu et al. 2008, 2012). In the case of a single Gaussian or Lorentzian shape, the line is allowed to be shifted with respect to Fe II. In this way the number of free parameters in these fits is the same as the number of free parameters in Edgeworth and Gauss-Hermite expansions.

The underlying broad Fe~II UV pseudo-continuum is modeled with the use of several theoretical and observational templates. Observational templates come from Vestergaard \& Wilkes (2001) and 
Tsuzuki et al. (2006), and they were derived for
the extreme case NLS1 object I Zw 1. The collection of theoretical templates calculated for different values of the density, 
turbulent velocity, and ionization parameter $\Phi$ comes from Bruhweiler \& Verner (2008).
 Finally, we also tried the observational template derived by  Hryniewicz et al. (2014) as a by-product of the analysis 
of the LBQS 2113-4538. The fitting of the components was performed with our own Fortran code, and the fit quality was determined at the basis of $\chi^2$ values. Templates are numbered in the following way: temp. 1 is the observational template from Vestergaard \& Wilkes (2001), temp. 2 is the observational
template of Tsuzuki et al. (2006), templates 3 -- 15 are theoretical templates of Bruhweiler \& Verner (2008), designated as d11-m20-20.5-735, d11-m30-20-5-735, d11-m20-21-735,
d10-5-m20-20-5, d11-m05-20-5, d11-m10-20-5, d11-m20-20-5, d11m30-20-5, d11-m50-20-5, d11-5-m20-20-5, d12-m20-20-5, d11-m20-20,d11-m20-21 in their paper, and temp. 16 is the
observational template from Hryniewicz et al. (2014).

Broad  band data fitting is done for the Novikov-Thorne accretion disk model (Novikov \& Thorne 1973). Model implementation, described in Czerny et al. (2011), includes all relativistic effects both in the disk structure and in light propagation to the observer. We neglect the outflows since they are expected to be unimportant in the case of high black hole mass (Laor \& Davis 2014).

\section{Results}

We analyzed ten high-quality RSS spectra of the quasar HE 0435-4312 obtained with the SALT telescope and covering three years. We now  model in detail the relatively narrow spectral range, 
2700 - 2900 \AA ~ in the rest frame of the object, assessing the variability of line intensity and shape for Mg II as well as underlying Fe II. We then discuss the physical conditions in the formation site of these two components, combining the spectral variability with constraints from the black hole mass determination and photoionization calculations using the broad band data.

The Mg II line variability is clearly seen in our data set, as illustrated in Fig.~\ref{fig:widma_sample}, where we plot a few examples of spectra renormalized at 2700 \AA. The change in the line equivalent width (EW) is clearly the dominant factor but visual inspection shows some change in the line shape. The line seems to shift towards longer wavelengths, but also the line peak is distorted. However, the underlying power law also varies, as indicated by the slope change at longer wavelengths. Also the underlying Fe II seems to change, judging from the visibility of the Fe II local bump at $\sim 2750$ \AA. Due to this complex behavior, quantitative statements require non-parametric studies and parametric spectral fitting.

\subsection{Non-parametric tests of the Mg II line evolution}
\label{sect:nonparam}
%______________________________________________ 
%
    \begin{figure}[ht]
    \centering
    \includegraphics[width=0.49\textwidth]{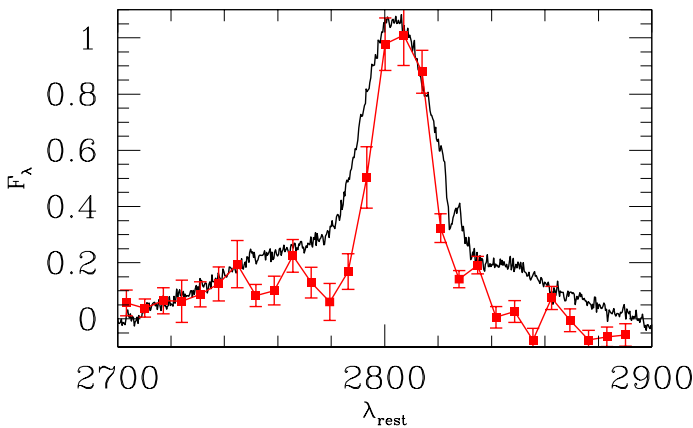}
    \caption{Mean (black line) and rms spectrum (binned and multiplied by a factor of 20; red line and points) of the  quasar HE 0435-4312 without Observation 6.}
               \label{fig:rms}%
    \end{figure}
%
%___________________________

Our spectra do not have an absolute calibration which makes the calculation of the mean and rms spectra more difficult. We thus calculated the mean spectrum as a simple average of the individual spectra so the spectra with a higher instrumental flux have a relatively higher weight. Since the exposure times and the telescope setup were similar, the differences in the instrumental fluxes between the spectra are not large. 

To see the character of the variability of the Mg II line, we calculated rms spectra (see Peterson et al. 2004) by renormalizing first all the individual spectra in the wavelength range 2700 - 2717 \AA, that is, in the blue part of the spectrum. The result is shown in Fig.~\ref{fig:rms}.  The mean shape of the line shows asymmetry, but it is partially due to the presence of  the underlying Fe II components. The rms spectra in the same units have very low normalization: the variations at the line peak position are of the order of 5 \%. Thus we multiply the rms by a factor of 20 for better visibility. The level of variability in the red part of the spectrum indicates variability in the slope of the underlying power law. The variation in Fe II is not clearly seen due to the performed renormalization. The mean and rms shapes of the Mg II line are basically similar (amplitude of the line varies the most) but variations are slightly enhanced in the red part of the line peak. Overall, the rms profile seems narrower than the mean profile, which is an opposite trend to that seen in the H$\beta$ line in  most of the reverberation-measured AGN (e.g., Peterson et al. 2004). However, the quality of our rms spectrum is rather low due to the way we perform the renormalization, and due to the low amplitude of the source variability. 

\begin{figure}
   \centering
%   \vskip - 2.0 true cm
   \includegraphics[width=0.45\textwidth]{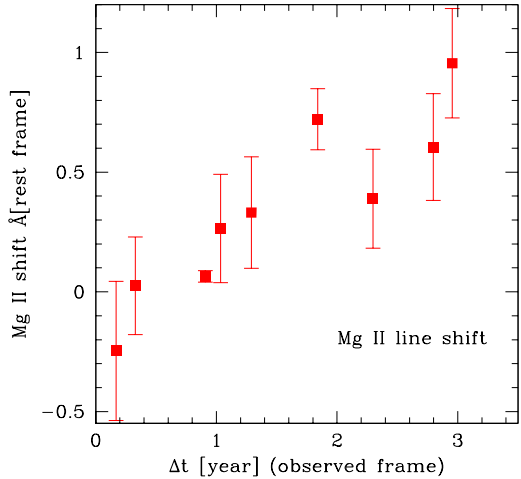}
   \caption{Shift  of the Mg II line in Observations 1 to 9 with respect to Observation 10, calculated directly from the data (see Sect.~\ref{sect:nonparam}). }
              \label{fig:line_evol}%
    \end{figure}

Since the most characteristic pattern seen in many quasar emission lines is the change of the line position with time (Liu et al. 2014), we determined the possible time-dependent line shift directly from the data. The application of the Interpolated Cross-Correlation Function (ICCF) method did not give a well specified maximum which would show the line shift amplitude. We thus tried the $\chi^2$ method for obtaining the line shift since we found this method more efficient for the time delay measurement (Czerny et al. 2013), and it was also used by Liu et al. (2014) in their two-epoch approach to the quasar line shift. With this method we see the overall line shift. We followed the method of Liu et al. (2014), that is, we used the shift in the pixel space, and determined the best shift value more accurately by fitting the resulting $\chi^2$ distribution.  The measured difference between the line position in Observations 1 and 10 was only $3.42^{+0.23}_{-0.18}$ in pixel units (one pixel corresponds to 0.6211 \AA~ in the observed frame) but it is determined reliably. In addition, the time evolution reveals a systematic trend (see Fig.~\ref{fig:line_evol}). These computations were performed selecting 100 data points in the line region (2780 - 2820 in the rest fame), but broadening somewhat the wavelength range does not change the conclusions. The trend, if interpreted formally as a linear shift, implies a change of the line position by $0.97 \pm 0.13$ \AA/year, or an acceleration of $104 \pm 14$ km s$^{-1}$ year$^{-1}$ in the quasar rest frame, assuming a redshift $z = 1.2231$.  The values measured by Liu et al. (2014) span the range from 10 to 200 km s$^{-1}$ year$^{-1}$, and only three quasars out of the 50 in their sample show larger acceleration than HE 0435-4312, so the change detected in our quasar is definitively one of the fastest.

%______________________________________________ 

\subsection{Individual spectra: single component fits}

We started the modeling of the Mg II doublet in the form of a single Gaussian and a single Lorentzian shape. In Table~\ref{tab:free_redshift} we give the fitted redshifts and the $\chi^2$ values for a fixed doublet ratio 1:1. The vast majority of observations, both for the Gaussian and Lorentzian fit, were best represented with the Fe II template 3. The Gaussian shape represents the data significantly better. The favored Gaussian broadening of the template was 900 km s$^{-1}$, only two observations required a slightly higher value, 1150 km s$^{-1}$. The mean redshift in all the data is $1.22230 \pm 0.00016$. We thus assumed this redshift value, fixed the Fe II broadening in all spectra at 900 km s$^{-1}$, and refitted the data using the Gaussian profile in order to determine the evolution of the Mg II and Fe II line.

The EW of the MG~II line line clearly changes, with the overall trend corresponding to the rise rate of $2.38 \pm 0.20$ \AA/year in the rest frame,
although the observed change is not strictly monotonic. The average value of EW(Mg~II) (20.7 \AA) locates the quasar among the typical sources, although on the lower side (e.g., Elvis et al. 2012; Dietrich et al. 2002 show the typical values of $\sim 40$ \AA, and in Vanden Berk et al. 2001 the composite spectrum has 32.3 \AA). The mean value of EW(Mg~II) in the DR7 quasar catalog of Shen et al. (2011) is 37.2 \AA~ (for sources in the redshift range 1.0 - 1.4, and with the measurement error smaller than 20\%). Still, it is well above the limit of 11 \AA~ defining Weak Line Quasars (e.g., Meusinger \&  Balafkan 2014). The line shape is similar to Mg II composites created by Tammour et al. (2016) at the basis of the Paris et al. (2014) quasar catalog.  

EW(Fe II) is also rising, so the increase in the EW(Mg~II) is not an artifact of the spectral decomposition into the Mg II and the underlying pseudo-continuum. The errors in the determination of Fe II properties are generally higher since such a broad feature is coupled more strongly to the determination of the underlying power law. However, the overall rise is well visible: $3.22 \pm 0.82$ \AA/year in the rest frame if fitted as a linear trend. This rise, in terms of a percentage change, is the same as for Mg II.  

\begin{figure}
   \centering
%   \vskip - 2.0 true cm
   \includegraphics[width=0.45\textwidth]{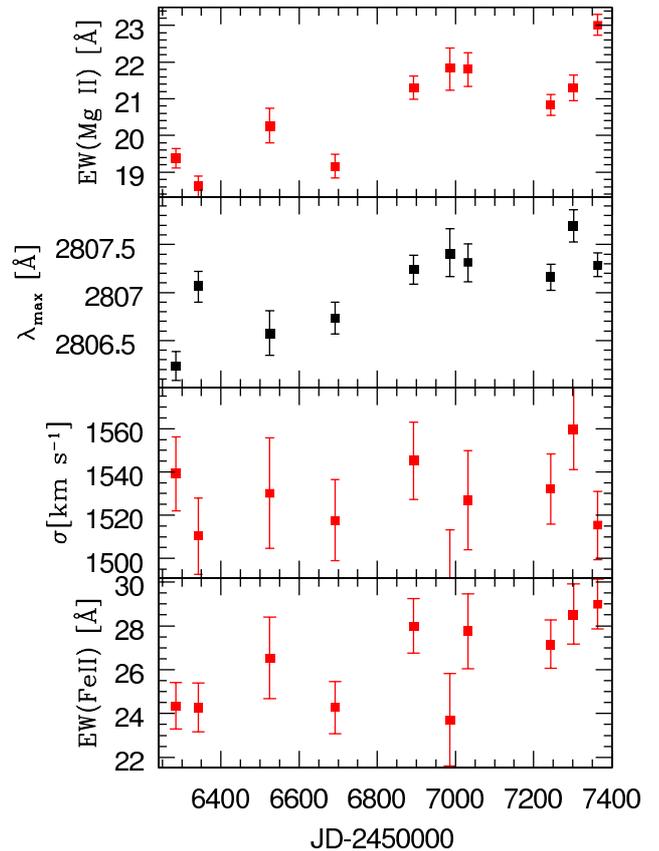}
   \caption{Evolution of the Mg II and Fe II lines with time: the total equivalent width, the position of the line maximum, and the line dispersion from single Gaussian fits: source redshift $z = 1.2223$, Fe II pseudo-continuum modeled with the template 3, 900 km s$^{-1}$ broadening. Errors represent 90 \% confidence level.}
              \label{fig:SG_trend}%
    \end{figure}

The line kinematic width does not show any systematic trend in time, with the mean value $\sigma = 1525 \pm 6$ km s$^{-1}$ (FWHM - full width at half maximum -  $3595 \pm 15$ km s$^{-1}$). However, the line position, represented by the position of the maximum, shows overall systematic rise  by $0.64 \pm 0.11$ \AA/year. This value is slightly lower than the value $0.97 \pm 0.13$ \AA/year, obtained directly from the non-parametric analysis. The discrepancy likely reflects some intrinsic change in the line shape, which enters in a different way in these two determinations. The fits are given in Table~\ref{tab:SG} and illustrated in Fig.~\ref{fig:SG_trend}.

Those results were obtained for a fixed doublet ratio 1:1. We thus checked how this assumption influences the fits, and in particular the line position and the spectrum decomposition into Mg II and Fe II. However, since the line position is a free parameter (specified by the shift $s$), and the line is broad enough that the doublet is unresolved, the change of the doublet ratio does not lead to any significant change in the EW(Mg II) or EW(Fe II). For example, in Observation 4 the assumption of the doublet ratio 2:1 leads to a decrease of the total $\chi^2$ by 2.0, line widths are unchanged, and the modification of the doublet ratio is fully compensated by the change in the line shift (parameter $s$ increased from $2.37 \times 10^{-3}$ to $2.81 \times 10^{-3}$), thus leaving the line peak also unchanged.

%______________________________________________ 

\subsection{Individual spectra: complex component fits}

%______________________________________________ 
%
    \begin{figure}[ht]
    \centering
    %   \vskip - 0.5 true cm
    \includegraphics[width=0.45\textwidth]{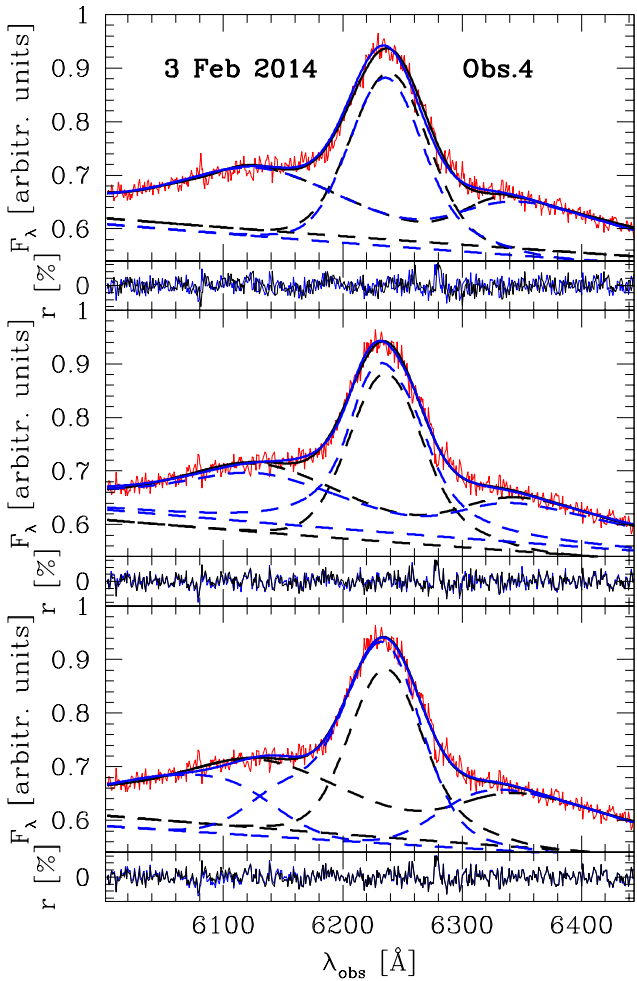}
    %\vskip - 1.5 true cm
    \caption{Comparison of the best fits of [upper panel:] a single Gaussian shape (black lines) and an Edgeworth shape (blue lines); [middle panel:] Edgeworth (black lines) and double Lorentzian shape (blue lines); [lower panel:] Edgeworth (black lines) and Gauss-Hermite shape (blue lines) for Observation 4 plotted in the observed frame. Continuous lines show the total model spectrum, dashed lines show the separate components.  Differences are relatively small in the upper two panels but the difference in $\chi^2$ of the fit is significant (see Table~\ref{tab:free_redshift}); the Gauss-Hermite expansion gives a different decomposition since it requires a different Fe II template (see text).}
               \label{fig:jedno_widmo_obs}%
    \end{figure}

A single Gaussian fit provides the measure of the line properties but leaves a limited possibility of more advanced interpretation. Therefore, we consider fits containing two more free parameters in order to check whether they offer a statistical improvement and better insight into the character of the observed variations. We test the Gauss-Hermite expansion, Edgeworth expansion, double Gauss, and double Lorentzian model, assuming a fixed doublet ratio 1:1. In preliminary fits the redshift has been treated as a free parameter in individual spectra. The derived redshifts and the fit quality are given in Table~\ref{tab:free_redshift}. 

The Edgeworth expansion and the double Lorentz (LL) model gave solutions similar in character to the single Gaussian fits, but with $\chi^2$ significantly improved. These solutions generally favored template 3 for the Fe II pseudo-continuum, and the optimum broadening of the Fe II template for the Edgeworth (E) and double Lorentzian (LL) model, was 1550 km s$^{-1}$.

An example of the fit using a single Gaussian and a Gauss-Hermite expansion is shown in Fig.~\ref{fig:jedno_widmo_obs}. The plot had to be made in the observed frame since the best fit redshifts in those two fits are different. We selected Observation 4 to make the plot since this observation had the lowest sky background and one of the highest values of the S/N. The difference between the best fit curves is not large, mostly seen in the peak and in the lowest parts of the wings. The difference in the EW of the Mg II and Fe II lines between the single Gauss fit and Gauss-Hermite fit is also not large: 19.08 \AA~ vs. 19.41 \AA~ for Mg II,  and 23.75 \AA~ vs. 22.31 \AA, respectively.

\begin{table}
\caption{Values of EW(Mg II) and EW(Fe II ) with errors (90 \% confidence level) for  template 3 and two Lorentzian component shape.}   % title of Table
\label{tab:final_LL}      % is used to refer this table in the text
\centering                          % used for centering table
\begin{tabular}{l r r r r r r r r r}        % centered columns (4 columns)
\hline\hline      % inserts double horizontal lines
Obs.    & Mg II     &  Fe II          \\
        & EW        &  EW             \\
        & \AA       &  \AA            \\
\hline                        % inserts single horizontal line
1       &  $22.67^{+0.30}_{-0.36}$ & $21.16^{+1.08}_{-1.17}$ \\
2       &  $21.79^{+0.13}_{-0.13}$ & $20.76^{+1.37}_{-1.23}$ \\
3       &  $23.50^{+0.50}_{-0.51}$ & $22.11^{+2.02}_{-2.07}$ \\
4       &  $22.67^{+0.38}_{-0.35}$ & $21.28^{+1.44}_{-1.35}$ \\
5       &  $25.58^{+0.44}_{-0.38}$ & $25.12^{+1.12}_{-1.13}$  \\
6       & $25.46^{+1.20}_{-1.01}$  & $18.91^{+2.56}_{-2.36}$ \\
7       &  $25.88^{+0.49}_{-0.56}$ & $23.90^{+1.94}_{-1.87}$ \\
8       &  $24.80^{+0.32}_{-0.38}$ & $22.57^{+0.50}_{-1.17}$ \\
9       &  $25.13^{+0.41}_{-0.44}$ & $25.19^{+1.39}_{-1.43}$ \\
10      &  $27.48^{+0.36}_{-0.39}$ & $24.49^{+1.66}_{-1.50}$ \\
\hline                                   %inserts single line

\end{tabular}
\end{table}

It is interesting to note that the dispersion in the derived
values of the redshifts from a Gauss-Hermite profile (0.00006)
is smaller by a factor of two than in the case of a single Gauss
or a single Lorentzian (both 0.00016). This already suggests
that the line shape varies, 
and the change in the line shape can be accommodated by the change in the $h_3$ and $h_4$ parameters of the Gauss-Hermite expansion (see Appendix for parameter descriptions). The small dispersion in the redshift from the Gauss-Hermite expansion is actually comparable to the systematic shifts of the order of 0.2 \AA~ (or, equivalently, 0.00004 in redshift) which we introduced during the recalibration of the data with sky line. 
In the further analysis we assume the same  redshift for all observations within the frame of a given line shape, and we refit the results. The mean redshift for all more complex solutions is 1.22310 (as measured with respect to the Fe II component). 
Selected parameters for these solutions are given in Tables~\ref{tab:final_Hermit} and~\ref{tab:final_LL}.

The parameters of the Edgeworth expansion show clearly that the line is not strictly Gaussian: the mean value of the skewness is $0.37 \pm 0.04 $, and the mean value of the kurtosis is $1.05 \pm 0.09$. Nevertheless, the obtained values of EW(Mg II) are similar to the values from a single Gaussian fit, and the rising trend is also very similar. The level of Fe~II contribution is systematically lower by some 10\%. The line peak again shows a systematic trend with a similar rate, $0.64 \pm 0.21$ \AA/year in the rest frame. We formally see also the change in the line shape parameters: the linear fit to the evolution of skewness gives the rise rate per year by $0.08 \pm 0.06 $ (in the rest frame), and $-0.17 \pm 0.15$  for kurtosis, but taking into account that the given errors are just statistical errors we cannot claim a firm detection in line shape, apart from the evolution of line intensity and position.

The fits based on the Gauss-Hermite expansion were more complex since for some observations template 3 was favored, as before, but for four observations template 1 was better. The difference between the fit quality as measured with total  $\chi^2$  was larger than 10.0 in five data sets, including an exceptionally large difference of 84 for observation 2. In order to measure the line evolution, we cannot mix fits with different templates since they lead to very different results. Thus, for better illustration, we used template 1 consequently in all the GH fits, and we refitted the spectra again assuming the average redshift. The fits are shown in Table~\ref{tab:final_Hermit}. 

The mean value of EW(Mg~II) in such fits (32.98 \AA) is much larger than in fits based on template 3. However, the time trend, if expressed in percentage, is practically the same, ($12 \pm 2$ per cent per year in the rest frame, single Gaussian model gave $16 \pm 3$ in the same units) so the problem in the spectrum decomposition does not affect the conclusion about the line change. Line shift estimates from complex fits give larger errors but the results are consistent with the simple Gaussian fit. 
The parameters measuring the line departure from Gaussianity do not show any systematic trends in those fits.

\begin{figure*}
   \centering
%   \vskip - 2.0 true cm
%   \includegraphics[width=0.95\textwidth]{razem.eps}
%\includegraphics[width=0.9\textwidth]{razem_ladne.eps}
\includegraphics[width=0.9\textwidth]{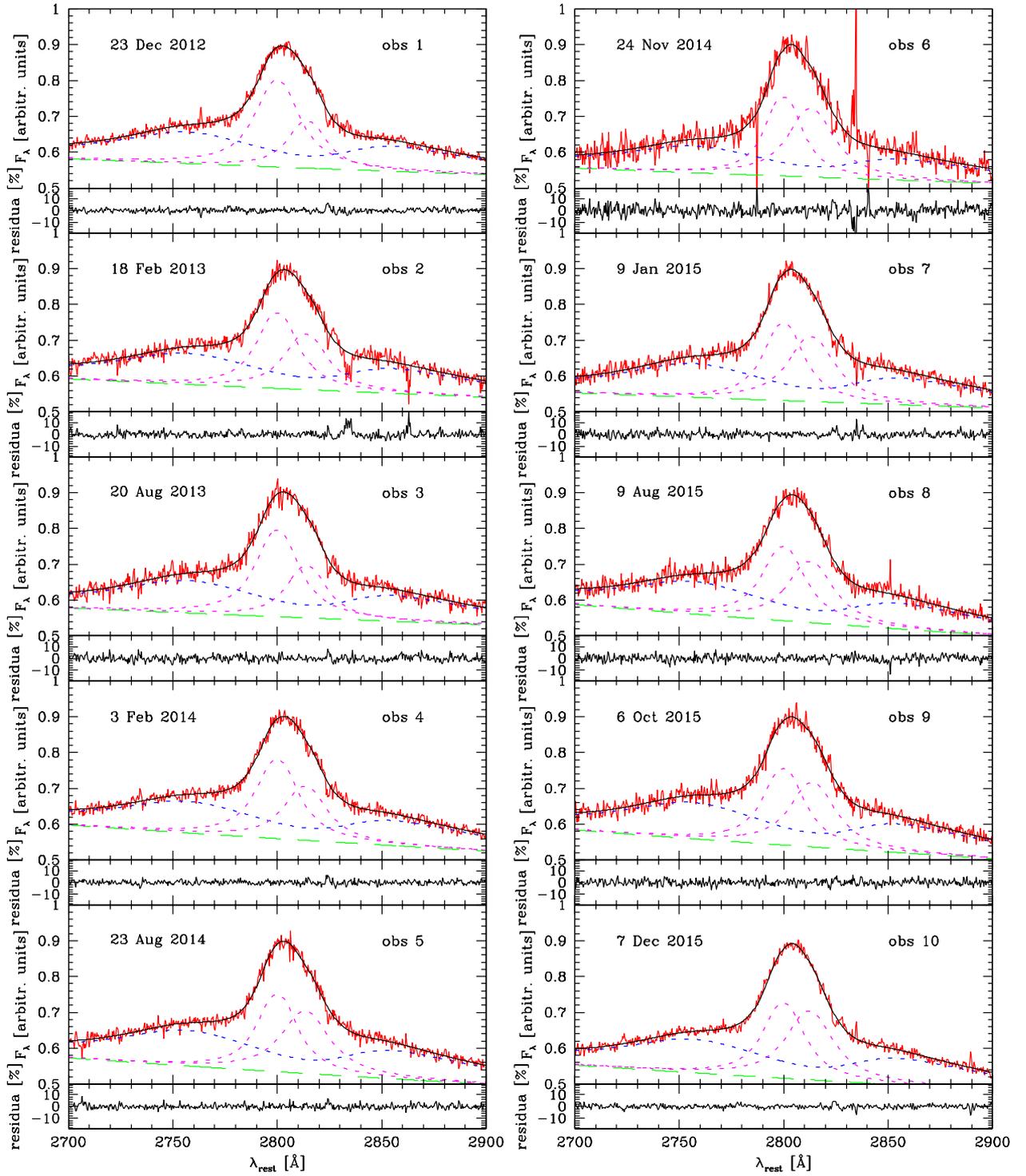}
   \caption{Best fit and residuals for ten observations, double Lorentzian fit; continuous lines show the model (red) and the data (black), 
   dashed green lines give the underlying power law, the dotted blue line represents the kinematically blurred Fe II pseudo-continuum, and dotted magenta lines mark the two 
   kinematic components of the Mg II. The position of the left component corresponds to rest frame for the adopted redshift, and is consistent with the position of Fe II. 
%\footnote{Spectra shown in Figs.~3 are available in electronic form at the CDS via anonymous ftp to cdsarc.u-strasbg.fr (130.79.128.5) or via http://cdsweb.u-strasbg.fr/cgi-bin/qcat?J/A+A/}. 
}
 \label{fig:widma}
    \end{figure*}

Fits with the use of two Lorentzian components formally gave the best fit for nine out of ten data sets, with the same number of free parameters as the other complex models.
The mean redshift for the double Lorentzian fit is the same
as for the Gauss-Hermite expansion; the redshift dispersion is
only slightly higher but still low, 0.00008.  The results are given in Table~\ref{tab:final_LL}. Lorentzian components are unresolved, their individual FWHM varies between 2000 and 2600 km s$^{-1}$, but they sum up to the total FWHM of the line 3576 km s$^{-1}$, and this value does not vary with time. Therefore, the two Lorentzians do not seem to have any specific interpretation other than providing lower $\chi^2$ than other solutions. All the spectra with corresponding fits are shown in Fig.~\ref{fig:widma}. We see again the same systematic shift of the line peak ($0.96 \pm 0.17 \AA$/year in the rest frame), the increase in the EW(Mg II) and EW(Fe II).

\subsubsection{Fe II templates}

We analyzed various Fe II templates as their shape strongly affects the fit to the Mg II line shape. Template 3, designated as d11-m20-20.5-735.dat in Bruhweiler \& Verner (2008), provides the best fit to Observation 4, and it is usually the best one for other observations and other choices of the Mg II line shape. This template 
was constructed assuming the density $10^{11}$ cm$^{-3}$, the turbulent velocity 20 km s$^{-1}$, and the ionization parameter $\log \Phi = 20.5$ cm$^{-2}$ s$^{-1}$. It is exactly the same template which was favored for LBQS 2113-4538 and 
I Zw 1 (Hryniewicz et al. 2014; Bruhweiler \& Verner 2008).

Template 9 with the same physical parameters but a different number of Fe II transitions included also provided a rather good fit. The change of the density towards a higher value,  $10^{12}$ cm$^{-3}$, gave the next optimum solution, and the changes in other plasma parameters gave total $\chi^2$ values worse by 25 or more than for template 3. 

The Mg II line is considerably shifted with respect to the Fe II pseudo-continuum: the mean value of the shift parameter $s$ in all ten observations fitted with template 3 ($2.4 \times 10^{-3}$) corresponds to the relative velocity 720 km s$^{-1}$, and it systematically increases with time. This indicates that the Fe II emission and Mg II emission do not come from the same region. Required broadening of the Fe II template (900 - 1550 km s$^{-1}$, depending on the chosen model of Mg~II) locates the Fe II region outside the Mg II emitting region, if the line width is interpreted as caused by Keplerian motion. Our result is consistent with the comparison of the width and shift of optical Fe II and H$\beta$ (Hu et al. 2008), and  reverberation measurement of an optical Fe II delay with respect to H$\beta$ (Chelouche et al. 2014). 

Gauss-Hermite expansion in several cases favored template 1, that is, the
 observational template based on I Zw I. This object was also modeled by Bruhweiler \& Verner (2008), and they found that template 3 models this object best, so the conclusions about the conditions in the Fe~II emitting region are also consistent. However, when modeling with template 1, we obtained a much lower mean shift between Fe~II and Mg II, only $\sim 150$ km s$^{-1}$, although the time trend in the line shift was the same. This may imply that the similar basic shift is also present in I Zw I.

\subsubsection{Doublet ratio}

The choice of the doublet ratio in principle influences the line position. We assume the same source redshift as before, $z = 1.22310$, but the doublet ratio is not constrained in more complex fits. For example, the best fit value for Observation 4, for the Edgeworth expansion, is achieved for the doublet ratio 2:1 but the corresponding change in total $\chi^2$ is only 1.6, so the improvement is not statistically significant. The Mg II line equivalent width and line skewness are practically unchanged, and the change in the doublet ratio was mostly compensated by the change in the line shift parameter ($s$ changed from $2.21 \times 10^{-3}$ to  $2.64 \times 10^{-3}$, i.e., more than 20 \%). The position of the line maximum changed only by 0.02 \AA, which is statistically insignificant. Therefore, we think that the line properties quoted for the fixed doublet ratio are representative.

%______________________________________________ 
%
%

%______________________________________________ 
   \begin{figure}
   \centering
%   \vskip - 2.0 true cm
   \includegraphics[width=0.49\textwidth]{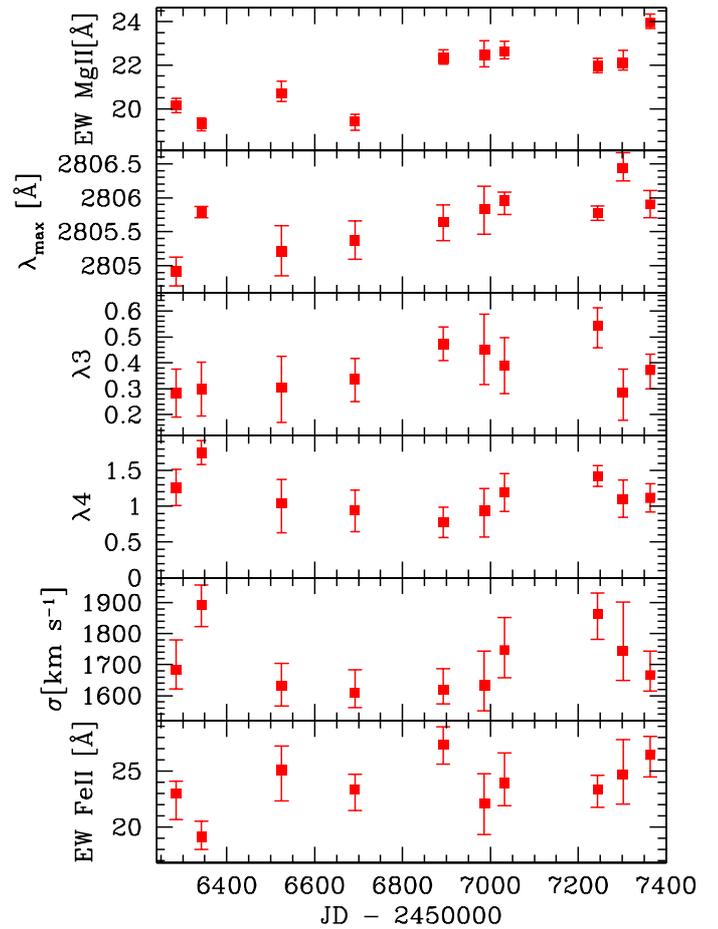}
   \caption{Evolution of the Mg II line and Fe II line with time for the Edgeworth expansion. Errors represent the 90 \% confidence level. }
              \label{fig:trend}%
    \end{figure}

\subsection{Comparison with Very Large Telescope (VLT) spectrum in the H$\beta$ range}
\label{sect:VLT}

%______________________________________________ 
   \begin{figure}
   \centering
%   \vskip - 2.0 true cm
%   \includegraphics[width=0.45\textwidth]{widmo_vel_OIII_new_ladne.eps}
 \includegraphics[width=0.45\textwidth]{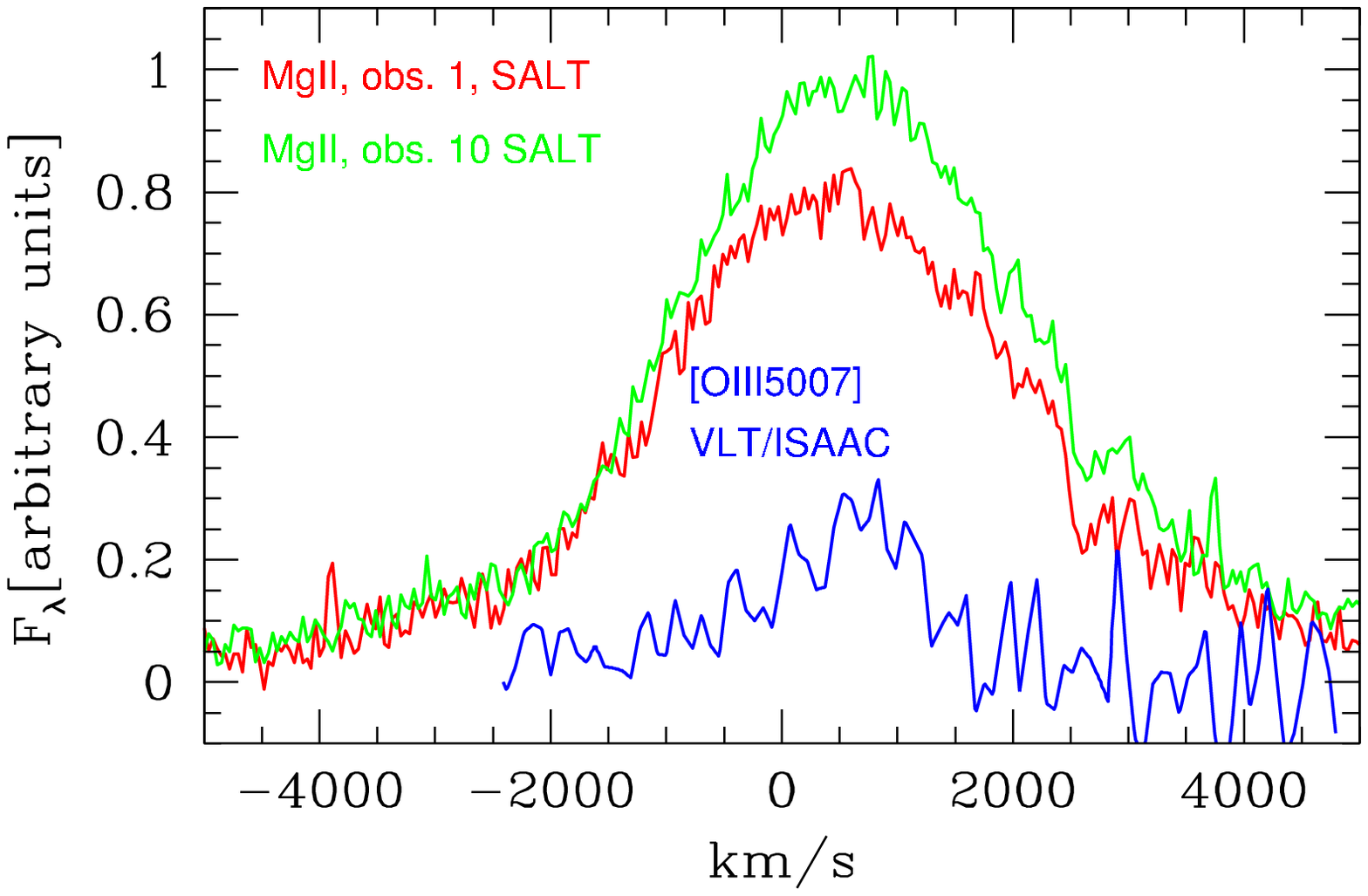}
   \caption{Comparison of the O[III]$\lambda$5007 line with the Mg II line in velocity space, with the underlying continuum subtracted; we used the nominal line positions, corrected for the difference in redshifts in data reduction between SALT and ISAAC/VLT; asymmetry of the O[III] line makes the qualitative comparison of the line positions difficult.}
              \label{fig:OIII}%
    \end{figure}
%
%
%______________________________________________ 
   \begin{figure}
   \centering
%   \vskip - 2.0 true cm
   \includegraphics[width=0.45\textwidth]{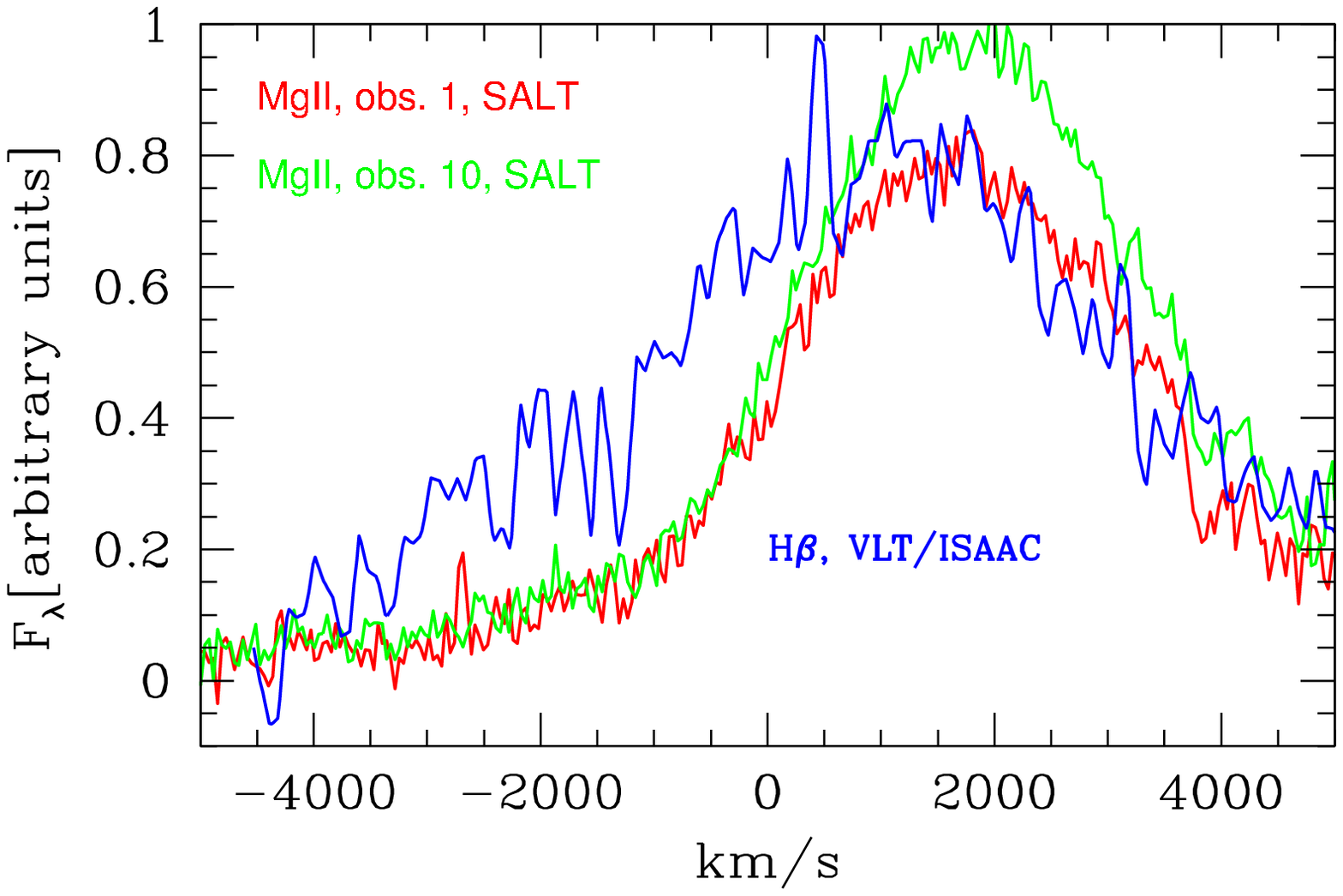}
   \caption{Comparison of the H$\beta$ line with the Mg II line in velocity space, with the underlying continuum subtracted; we used the nominal line positions, corrected for the difference in redshifts in data reduction between SALT and ISAAC/VLT; the Mg II line seems somewhat narrower, but the overall shape is similar to the shape of H$\beta$ in this source, and the difference in the position of the centroid is qualitatively in agreement with the extension of the long-term trend observed by us during the last three years.}
              \label{fig:Hbeta}%
    \end{figure}
Since our three-year long campaign does not clearly resolve the issue of the intrinsic change in the line shape, apart from the line systematic shift, we now include older observations which might give insight into the Mg II position in the rest frame and its long-term evolution.
The object was observed with VLT/ISAAC in the IR, on 17 September 2004, so the shape of the H$\beta$
line is available for this source (Marziani et al. 2009). The H$\beta$ line is single peaked close to the top, with FWHM of 5500 km s$^{-1}$ but the line as a whole is not exactly symmetric, with kurtosis index C.I. (for definition, see Marziani et a. 1996) of $0.38 \pm 0.08$; on the other hand, the asymmetry index A.I. introduced by Marziani et al. (1996) is small and consistent with zero  ($0.08 \pm 0.14$). This asymmetry index is specified at the 1/4 of the line height. For comparison, we also calculated the same asymmetry indices for the  Mg II line, using the double Lorentz decomposition since it fits the line shape best. The A.I. of Mg II varies between 0.05 and 0.11 in different observations (the largest values are in the middle of our campaign). The C.I. index varies between 0.42 and 0.44, in perfect agreement with the H$\beta$ shape and without any evolutionary trend.  
% see maksimum.dat 

The equivalent width of H$\beta$ is large, $118 \pm 24$ \AA, larger than typical quasar values (57.4 \AA, Forster et al. 2001) while the EW of the Mg II in this source ($ \sim 24.5$ \AA, see Table~\ref{tab:final_LL}) is somewhat lower than typical values (37.1 \AA, Forster et al. 2001). Larger kinematic width of H$\beta$ in comparison with Mg II (see values for a single Gaussian fit in Table~\ref{tab:SG} or Fig.~\ref{fig:trend}) is consistent with Eq.~6 of Wang et al. (2009), and the results of Marziani et al. (2013b), although Trakhtenbrot \& Netzer (2012) find no difference between Mg II and H$\beta$ width for objects with FWHM narrower than 6000 km s$^{-1}$. 

We overplotted examples of Mg II lines on O[III] and H$\beta$ in velocity space (see Figs.~\ref{fig:OIII} and \ref{fig:Hbeta}). The consistency seems satisfactory. The maximum of O[III] line roughly coincides with the current position of the Mg II line, but both are shifted with respect to the adopted source redshift. The red wing of H$\beta$ coincides with Mg~II in Observation 1, but the blue wing is more profound. Mg II in Observation 10 is shifted towards the red. Overall, it looks like a consistent slow evolution towards the red on timescales of 12 years. The line drift rate in the observed frame (47 km s$^{-1}$/year, see Sect.~\ref{sect:nonparam}) would accumulate to a shift of over 560 km s$^{-1}$, putting the centroid of Mg II and H$\beta$  roughly in agreement. However, this comparison should be treated with caution.

\subsection{black hole mass determination from the Mg II line and broad band data fits}
\label{sect:BHmass}

We test the single component character of the Mg II line by comparing the black hole mass obtained from the total FWHM to the black hole mass value from the disk fitting to the broad band continuum.
We use the general expression for the black hole mass determination
\begin{equation}
M = A \bigl({ \lambda L_{\lambda, 3000}\over 10^{44} {\rm erg~s}^{-1}}\bigr)^B \bigl({FWHM \over 1000 {\rm km~s}^{-1}}\bigr)^C,
\end{equation}
where the coefficients $A$, $B$, and $C$ depend on the specific formula ($5.6 \times 10^6 M_{\odot}$,0.62, 2.0 in Trakhtenbrot \& Netzer 2012, $3.4 \times 10^6 M_{\odot}$, 0.58, 2.0 in Kong et al. 2006, and
$1.3 \times 10^7 M_{\odot}$, 0.5, 1.51 in Wang et al. 2009), and give the black hole mass values of $2.2 \times 10^9 M_{\odot}$ (first method), and  $1.1 \times 10^9 M_{\odot}$ (the other two methods). For the FWHM we used 3576 km s$^{-1}$, the mean of the values from the double Lorentzian fits, and we assumed the logarithm of the
monochromatic flux of 46.396 at 5100 \AA. 

We now check whether these black hole mass values are consistent with the broad band spectrum of the source. The big blue bump in this quasar has clearly a disk-like shape and the maximum of the spectrum is well within the observed range, but modeling is far from unique. If we correct the data just for the Galactic extinction (see Sect.~\ref{broad_band_spectrum}) and assume a black hole mass of $2.2 \times 10^9 M_{\odot}$, we can represent well most of the broad band spectrum. We show the best fit to the optical/UV continuum in Fig.~\ref{fig:broad_band}. The best fit parameters are: the accretion rate, $\dot m$, is 0.58 in Eddington units, the  viewing angle is 23$^{\circ}$, and the spin is 0 (non-rotating black hole). The fit is not strictly unique since the normalization of the optical continuum constrains the accretion rate and the inclination while the extension of the spectrum to UV constrains the accretion rate and spin. However, a fast spinning black hole is excluded, and a moderately spinning black hole would require a still smaller viewing angle. The far-UV GALEX point in all cases is not well represented (see Fig.~\ref{fig:broad_band}), but this is a frequent problem in quasar fitting (e.g., Fig.~17 in Koratkar \& Blaes 1999 for a fit to the composite spectrum). It likely implies some form of intrinsic extinction. In the case of a single object, variability may also cause such an effect (the data points are not simultaneous). If we use a smaller black hole mass, $\sim 10^9  M_{\odot}$ or less, appropriate if any of the two Lorentzians should be used to determine the mass, we definitely overproduce the far UV part even if a counter-rotating black hole is allowed. Thus, this supports the conclusion that apart from the line departure from the Gaussian shape, the line indeed forms a single component and the full value of the FWHM should be used for the black hole mass measurement.

\subsection{Local conditions and the covering factor in the BLR}

%______________________________________________ 
   \begin{figure}[!ht]
   \centering
   \includegraphics[width=0.35\textwidth]{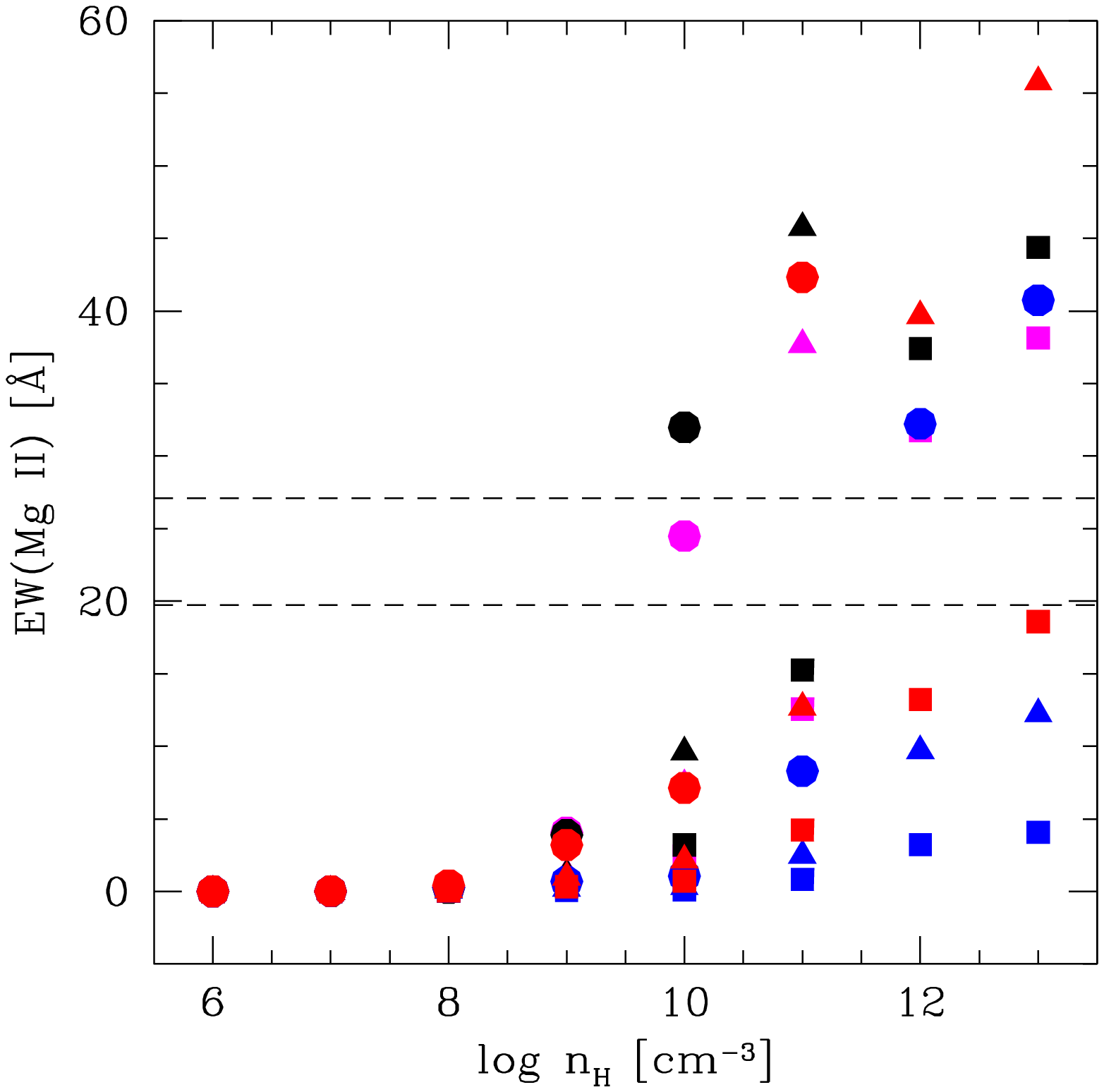}
  \includegraphics[width=0.35\textwidth]{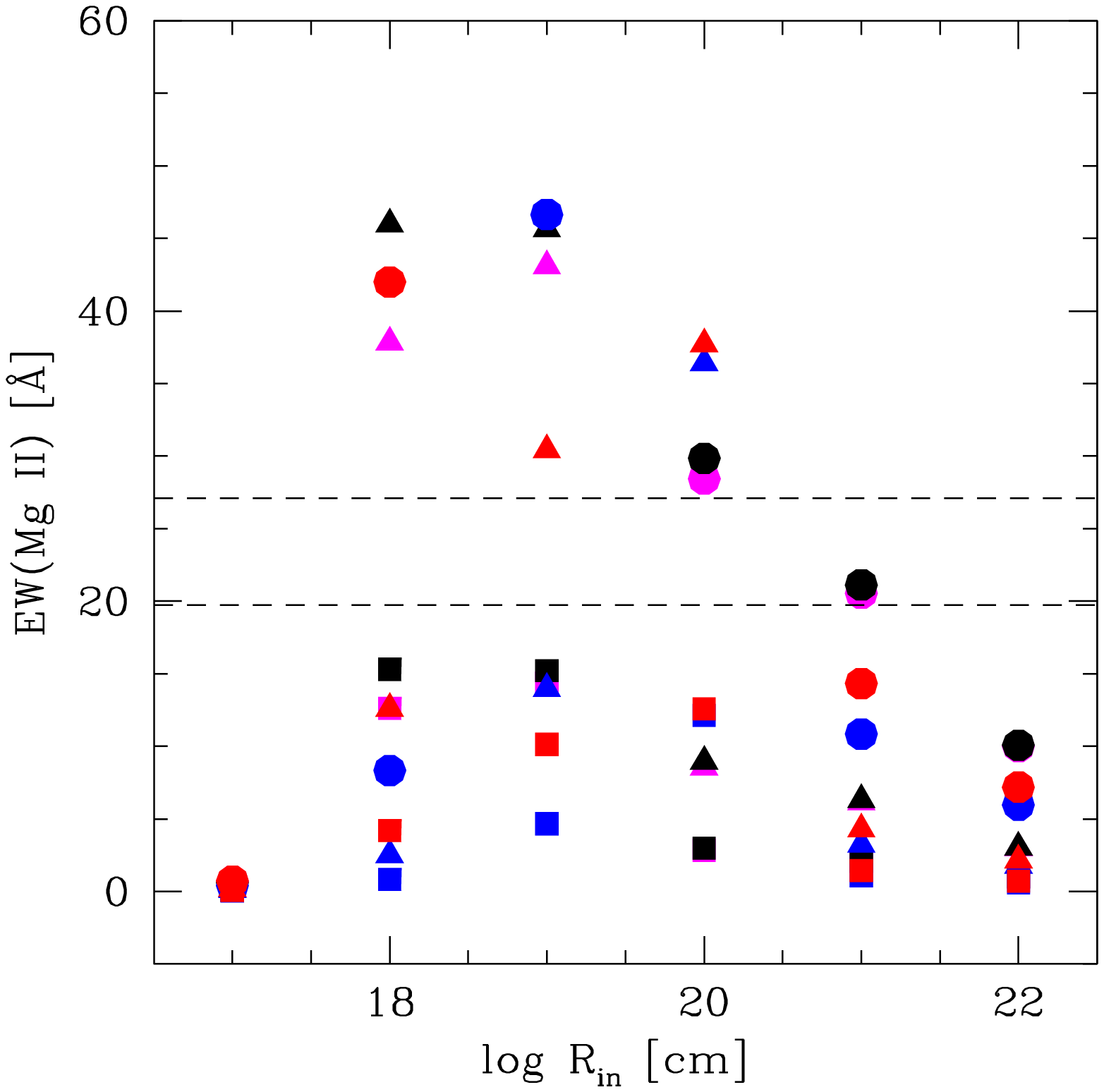}
  \includegraphics[width=0.35\textwidth]{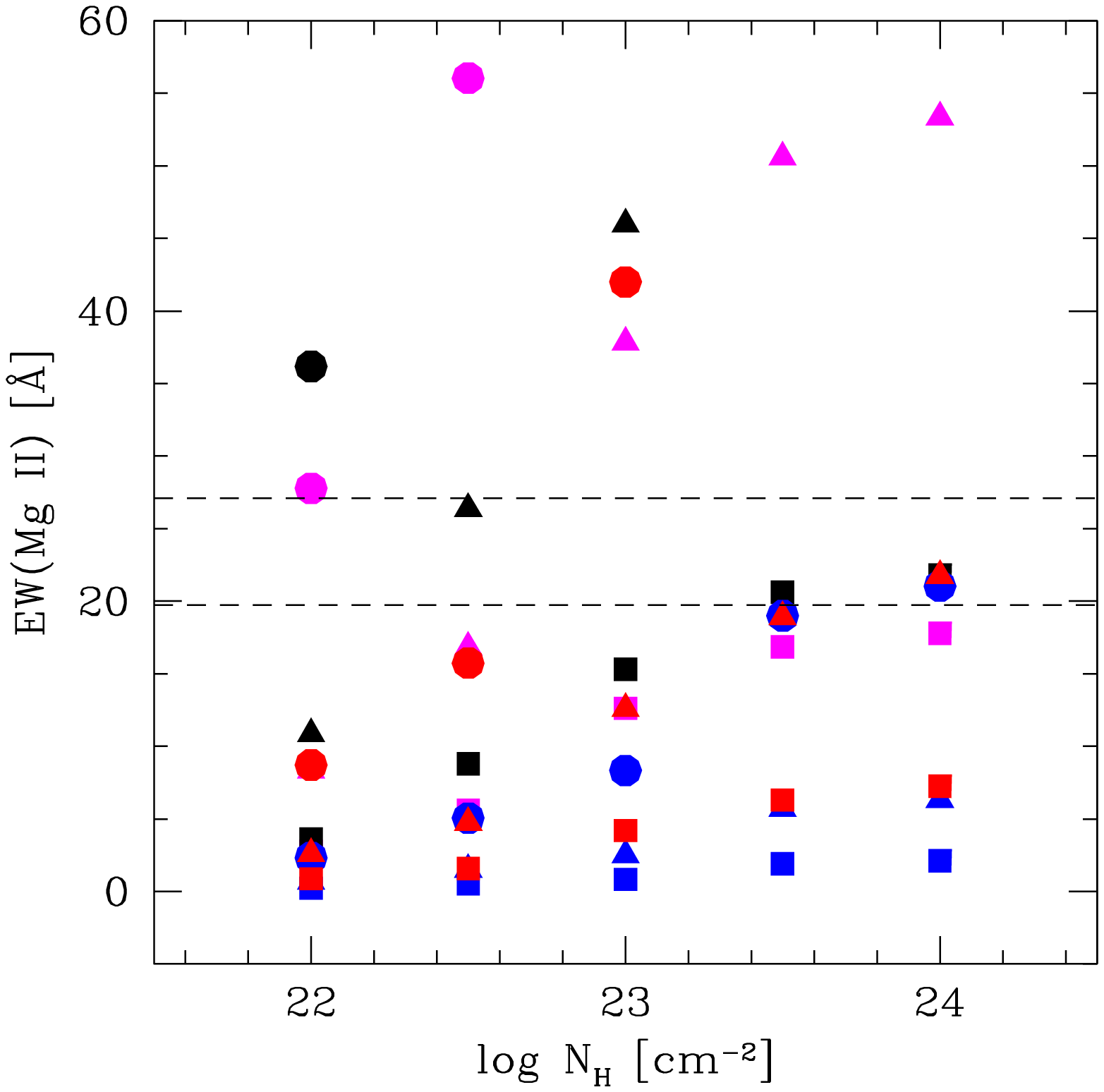}
   \caption{Dependence between EW(Mg II) and local cloud density, $n_H$, BLR radius, $R_{in}$, and the hydrogen column, $N_H$, with respect to the fiducial values $10^{11}$ cm$^{-3}$, $10^{18}$ cm, and $10^{23}$ cm$^{-2}$ obtained with CLOUDY for the broad band continuum from Fig.~\ref{fig:broad_band}, for four values of the turbulent velocity (10 km s$^{-1}$, blue, 100 km s$^{-1}$, red, 500 km s$^{-1}$, magenta, 1000 km s$^{-1}$, black), and three values of the covering factor (1.0, circles; 0.3, triangles; 0.1, squares).  Dashed horizontal lines mark the minimum and the maximum value from Table~\ref{tab:SG}.}
              \label{fig:turbulences}%
    \end{figure}
    
The systematically increasing shift between the Fe II emission and Mg II emission clearly suggests that these two emitting regions are different. We thus analyze other properties of the emitting material to check whether the kinematic separation is also supported by the difference in the local conditions of the Mg II and Fe II emitting plasma.

The BLR clouds are ionized by the optical/UV photons from the thermal emission of the accretion disk. The construction of the broad band continuum (see Fig.~\ref{fig:broad_band}) allows us to model Mg II line production and to obtain some constraints for the BLR properties. We do not limit the incident flux to the disk emission but we also include the IR part of the spectrum. Since the accretion disk models well the optical/UV part, we think that the X-ray emission is unlikely to be strong in this source and we neglect it. The bolometric luminosity of the source, integrated over the IR/opt/UV part, is $\log L_{bol} = 46.94$, where $L_{bol}$ is expressed erg s$^{-1}$. 

To calculate the EW of the Mg II line and the BLR covering factor, $C_f$,  we use the radiation transfer code CLOUDY
(Ferland et al. 2013), version 13.03. We use a simple single cloud model, since we have just one emission line well measured, and the Fe II pseudo-continuum. We calculate a grid of solutions  with and without turbulence, for a range of cloud locations, $R_{in}$, hydrogen local densities, $n_H$, and hydrogen column densities, $N_H$. The fiducial parameter values adopted by us are $n_H = 10^{11}$ cm$^{-3}$, $R_{in} = 10^{18}$ cm, and $N_H = 10^{23}$ cm$^{-2}$. We consider the solutions for the covering factor
\begin{equation}
C_{f} = {{\Omega} \over {4\pi}} 
\end{equation}
of the order of $\sim 0.1 - 0.3$ as satisfactory (e.g., Gaskell et al. 2007; Peterson 2006 and references therein).

We first performed computations assuming the turbulent velocity of 10 and 100 km s$^{-1}$ (see Fig.~\ref{fig:turbulences}, blue and red symbols). However, the solutions with such low turbulent velocities underpredict the EW of Mg II, unless the covering factor is close to one, and they maximize the emissivity at too large distances, $10^{19} - 10^{20}$ cm$^{-2}$. Therefore, we considered the solutions with strong turbulence. As shown in Fig.~\ref{fig:turbulences}, in the case of the  turbulent velocities  $\sim 500$ km s$^{-1}$ we can reproduce well the line strength for a covering factor $\sim 0.2$. Such a high turbulent velocity is consistent with the turbulent velocity of  $ 400$ km s$^{-1}$ deduced for the H$\beta$ line in NGC 5548 based on modeling the line shape (Kollatschny \& Zetzl 2013). 

A large turbulent velocity is consistent with the Failed Radiatively Accelerated Dusty Outflow (FRADO) model of the BLR (Czerny et al. 2015), based on the idea of Czerny \& Hryniewicz (2011). In this case both outflow and inflow are expected at a given location, which can be approximated as turbulence. On the other hand, this kind of failed wind may cause additional mechanical heating of the BLR clouds, as postulated by, for example, Dumont et al. (1998). This was not included in our computations.

We compare these conditions with the requirements for the Fe II formation. Our best template implies a density of $10^{11}$ cm$^{-3}$, a logarithm of the ionization parameter $\Phi = 20.5$, and a turbulent velocity of 20 km s$^{-1}$. Such local density is consistent with the density required for efficient formation of the Mg II line. The ionization parameter allows us to test the distance of Fe II formation. Our broad band continuum implies a flux of the hydrogen ionizing photons of $4.6 \times 10^{57}$ photons s$^{-1}$, and this total flux agrees with the required local flux at $1.08 \times 10^{18}$ cm, that is, close to the expected location of the BLR. This is in agreement with the required broadening, corresponding to $FWHM = 2100$ km s$^{-1}$, locating the Fe II emission not much farther than the Mg II region.

The templates of Bruhweiler \& Verner (2008) are provided with the proper normalization with respect to the incident flux $\Phi$. This allows us to obtain the covering factor for Fe II clouds required to reproduce the level of the pseudo-continuum in the data. This covering factor is very small, only 0.01, much lower than acceptable for Mg II. 

Thus, the formation regions of Fe II and Mg II are widely different, despite the similarity in the distance and density, since the covering factor and the turbulence in the Mg II production region is high, and in the Fe II production region it is low, and there is a systematic shift between Mg II and Fe II increasing over the years. It might imply that the Fe II region is actually hidden behind the turbulent Mg II, and it is therefore much more quiet and less exposed to irradiation by the central source.    

We cannot compare our results with previous observational determinations of the Fe II to Mg II ratios (e.g., Dietrich et al. 2003) since they were based on broader band spectra and include the Fe II emission from the spectral range outside the 2700 - 2900 \AA~ band used in our analysis.

%______________________________________________________________

\section{Discussion}

We analyzed in detail ten observations of the Mg II$\lambda 2800$ line region in the quasar HE 0435-4312, covering three years. The equivalent width of the Mg II line as well as the Fe II pseudo-continuum rise with time. This is consistent with the rise in the continuum. What is more, we observe a systematically increasing shift between Mg~II and Fe II over the years, with the average rate corresponding to the acceleration by $104 \pm 14$ km s$^{-1}$ year$^{-1}$ in the quasar rest frame. Shifts between quasar emission lines have been detected before in a number of sources by comparing two-epoch spectra (Shen et al. 2013, Liu et al. 2014). Our observation supports the interpretation of such shifts as systematic trends, since our ten observations allow us to follow the line evolution, and the observed acceleration is one of the largest in comparison with the results from the Liu et al. (2014) sample. Despite the use of a number of line decompositions, we do not clearly detect any intrinsic change in the line shape. Marginal evolution of the kurtosis in the Edgenworth expansion is not convincing. On the other hand, the Mg II line shape shows a significant departure from the Gaussian shape, with both skewness, but mostly kurtosis, different from zero. The line is best modeled using two Lorentzian components but otherwise there is no indication of any two-component character of the line. The black hole mass estimate from the total line kinematic width and from the fitting of the broad band spectra supports this view.

There is no evidence for the systematic evolution of the Fe II itself, and the conditions in the Mg II and Fe II emitting regions are considerably different. Modeling with CLOUDY implies that the Mg II region has a large covering factor and a very large local turbulence velocity, of the order of $\sim 500$ km s$^{-1}$, while the Fe II region shows a small covering factor and a low turbulence velocity, $\sim 20$ km s$^{-1}$, although the difference in the kinematic width does not imply a much larger distance of the Fe II formation in comparison to Mg II. 

Several mechanisms might in general explain the variability of the emission line shape in a given source: (i) binary black hole (e.g., Popovic et al. 2000, Bon et al. 2012, Shapovalova et al. 2016); (2) spiral waves (e.g., Storchi-Bergman et al. 1995, Karas et al. 2001);  (3) spots on the disk surface; (4) precession of an elliptical disk (Eracleous et al. 1995, Storchi-Bergman et al. 1995);  (5) tidal disruption events (Komossa et al. 2008);  (6) warped disk model (e.g., Halpern \& Eracleous 2000);  (7) accretion disk instabilities (Mineshige \& Shields 1990; Czerny et al. 2009); (8) disk/wind model (Chiang \& Murray 1996,  Sulentic et al. 2009, Flohic et al. 2012); (9) tilted disk - Bardeen-Petterson effect (e.g., Nealon et al. 2015). Most of these scenarios imply periodic changes. In our data we see at best a fraction of a period. If we consider the trend seen in SALT Mg II as a continuation of the net shift of the Mg II centroid in comparison to H$\beta$ measured in 2004 with ISAAC/VLT, then the trend of systematic line shift lasts already for 12 years. 

Since our observations are not long enough to cover the possible period, we can refer to another signature of the process which is the kinematic shift of the BLR with respect to the host galaxy. Such an effect is expected in the case of a binary black hole (Comerford et al. 2009), and likely in other scenarios as well. However, in our case the position of [OIII] is hard to estimate reliably due to the strong line asymmetry and low quality of the data in that part of the ISAAC/VLT spectrum. It is, however, remarkable that the Fe II pseudo-continuum, likely hiding behind the Mg II region, does not show any systematic evolution in the velocity space. This might suggest rather a local phenomenon related to the disk structure as the factor governing the Mg II line evolution.

Finally, the rate of the change in the Mg II position in this intermediate redshift quasar is truly remarkable. The acceleration measured by us, $104 \pm 14$ km s$^{-1}$ year$^{-1}$ in the quasar rest frame, can be compared to the acceleration in the radial velocity component of the Keplerian motion. This radial velocity is given by $v_r =\Omega_K R \sin i \sin (\Omega_K t + \phi)$, and the corresponding acceleration is
\begin{equation}
a_r = \Omega_K^2 R \sin i \cos (\Omega_K t + \phi),
\end{equation}
where $\Omega_K = G M/R^3$ is the Keplerian velocity, $R$ is the orbital radius, $i$ the inclination angle to the orbital plane, and $\phi$ is the phase of the motion. Assuming the unrealistic setup: inclination angle $i = 90$ deg, and optimum phase maximizing the acceleration $(\Omega_K t_o + \phi) = 0$ in the middle of our observing campaign, this expression reduces to
\begin{equation}
a_\mathrm{max} = {G M \over R^2}.
\end{equation}
Taking the numbers appropriate for the BLR of the quasar, $M = 2.2 \times 10^9 M_{\odot}$ and the BLR radius from the formula
\begin{equation}
\log R_{BLR} = 1.555  + 0.542  \log {\lambda L_{\lambda}^{5100 \AA} \over 10^{44} {\rm erg~ s}^{-1}} \, {\rm  [lt. ~days]}\;,
\end{equation}
as obtained for nearby objects (Bentz et al. 2013), for $\log \lambda L_{\lambda}^{5100 \AA} = 46.57$ we obtain the maximum expected acceleration of 0.055 cm s$^{-2}$, a factor of six lower than the values measured by us (0.33 cm s$^{-2}$, when expressed in the same units). This points out that the observed acceleration is unlikely to be related to Keplerian motion of any disk feature, or to black hole binarity. Instead, it rather points towards a disk complex structure (spiral waves, ellipticity) illuminated by slightly anisotropic variable emission from the central parts, for example the innermost tilted/precessing disk, where the timescales for change are much shorter. Such models were considered, for example, by Gaskell (2011). However, further monitoring is necessary to support this conclusion.

\begin{appendix}
\section*{Appendix A}
\label{Appendix}

Hermite polynomials are frequently applied to characterize small departures from Gaussianity. Two types of this expansion are in use, and they are not equivalent. Gauss-Hermite expansion is based directly on the system of orthonormal Hermite functions and the Edgenworth expansion is constructed as an expansion in distribution momenta. If limited to the first two terms, the two expansions read:
\begin{eqnarray}
F_{\mu}^{Gauss-Hermit}&=&{1 \over \sqrt{2 \pi} \sigma}e^{(-\mu^2/2)} \bigl[1 + {h_3 \over \sqrt{6}}(2 \sqrt{2}\mu^3 - 3 \sqrt{2} \mu) + \nonumber \\
& &  {h_4 \over \sqrt{24}} (4\mu^4 - 12 \mu^2 +3)\bigr],
\end{eqnarray} 
and
\begin{eqnarray}
F_{\mu}^{Edgenworth}&=&{1 \over \sqrt{2 \pi} \sigma}e^{(-\mu^2/2)} \bigl[1 + {\lambda_3 \over 6}(\mu^3 - 3 \mu) + \nonumber \\
& & {\lambda_4 \over 24} (\mu^4 - 6 \mu^2 +3)\bigr],
\end{eqnarray}
where
\begin{equation}
\mu = {\lambda - \lambda_0(1 + s) \over \lambda_0\sigma/c},
\end{equation}
and $\sigma$ is the dispersion measure and $c$ is the speed of light. In the case of the Edgenworth expansion, the two coefficients $\lambda_3$ and $\lambda_4$ are exactly equal to skewness and kurtosis, respectively, and this is why this expansion is broadly used in cosmological applications when testing the non-Gaussianity of the observed distributions (e.g., Bianchi et al. 2016). The interpretation of $h_3$ and $h_4$ is not direct, although they measure the skewness and kurtosis to some extent. On the other hand, the characteristic property of the Gauss-Hermite expansion is to easily model the Gaussian line with additional broad wings, while the Edgenworth expansion does not easily do that and this is the reason why the Gauss-Hermite expansion is broadly used in emission line fitting (see Bendo et al 2016 for a recent example). From the convergence point of view, the Edgeworth expansion is considered to be better (Blinnikov \& Moessner 1998) since it allows for error control.

\end{appendix}

\begin{acknowledgements}
We are deeply grateful to the referee for comments which allowed us to reshape the paper considerably. Part of this work was supported by Polish grants Nr. 719/N-SALT/2010/0, UMO-2012/07/B/ST9/04425, and 2015/17/B/ST9/03436/. The spectroscopic 
observations reported in this paper were obtained with the Southern African Large Telescope (SALT),  
proposals 2012-2-POL-003, 2013-1-POL-RSA-002,2013-2-POL-RSA-001, 2014-1-POL-RSA-001, 2014-2-SCI-004, 2015-1-SCI-006, 2015-2-SCI-017. BC acknowledges the financial support from the Fellowship granted by the Chinese Academy of Sciences. J\'{S}, BC, KH, MK, AK, and BY
acknowledge the support by the Foundation for Polish Science through the
Master/Mistrz program 3/2012.  
The Fe II theoretical templates described in 
Bruhweiler \& Verner (2008) were downloaded from the
web page \url{http://iacs.cua.edu/personnel/personal-verner-feii.cfm} with the permission of the authors.
Part of this work is based on archival data, software, or on-line services provided by the ASI Science Data Center (ASDC).
This research has made use of the USNOFS Image and Catalogue Archive
   operated by the United States Naval Observatory, Flagstaff Station
   (\url{http://www.nofs.navy.mil/data/fchpix/}).
This publication makes use of data products from the Wide-field Infrared Survey Explorer, which is a joint project of the University of
California, Los Angeles, and the Jet Propulsion Laboratory/California Institute of Technology, funded by the National Aeronautics and
Space Administration.

This publication makes use of data products from the Two Micron All Sky Survey, which is a joint project of the University of
Massachusetts and the Infrared Processing and Analysis Center/California Institute of Technology, funded by the National Aeronautics
and Space Administration and the National Science Foundation.
The CSS survey is funded by the National Aeronautics and Space
Administration under Grant No. NNG05GF22G issued through the Science
Mission Directorate Near-Earth Objects Observations Program.  The CRTS
survey is supported by the U.S.~National Science Foundation under
grants AST-0909182.
Based on observations made with the NASA Galaxy Evolution Explorer.
GALEX is operated for NASA by the California Institute of Technology under NASA contract NAS5-98034. 
This research has made use of the NASA/IPAC Extragalactic Database (NED) which is operated by the 
Jet Propulsion Laboratory, California Institute of Technology, under contract with the National Aeronautics and Space Administration. 
\end{acknowledgements}

\end{document}